\documentclass[
reprint,
amsmath,
amssymb,
aip,
superscriptaddress,
notitlepage,
nofootinbib,
]{revtex4-2}

\usepackage[colorlinks, breaklinks=true, linkcolor=blue, citecolor=blue, linktocpage=true]{hyperref}
\usepackage{color}
\usepackage{graphicx}
\usepackage{dcolumn}
\usepackage{bm}
\usepackage{comment}

\usepackage{txfonts}

\begin{document}

\title{Microwave-to-Optical Quantum Transduction of Photons for Quantum Interconnects}

\author{Akihiko Sekine}
\email{akihiko.sekine@fujitsu.com}
\affiliation{Fujitsu Research, Fujitsu Limited, Kawasaki 211-8588, Japan}
\author{Ryo Murakami}
\affiliation{Fujitsu Research, Fujitsu Limited, Kawasaki 211-8588, Japan}
\author{Yoshiyasu Doi}
\affiliation{Fujitsu Research, Fujitsu Limited, Kawasaki 211-8588, Japan}

\date{\today}

\begin{abstract}
The quantum transduction, or equivalently quantum frequency conversion, is vital for the realization of, e.g., quantum networks, distributed quantum computing, and quantum repeaters.
The microwave-to-optical quantum transduction is of particular interest in the field of superconducting quantum computing, since interconnecting dilution refrigerators is considered inevitable for realizing large-scale quantum computers with fault-tolerance.
In this review, we overview recent theoretical and experimental studies on the quantum transduction between microwave and optical photons.
We describe a generic theory for the quantum transduction employing the input-output formalism, from which the essential quantities characterizing the transduction, i.e., the expressions for the transduction efficiency, the added noise, and the transduction bandwidth are derived.
We review the major transduction methods that have been experimentally demonstrated, focusing on the transduction via the optomechanical effect, the electro-optic effect, the magneto-optic effect, and the atomic ensembles.
We also briefly review the recent experimental progress on the quantum transduction from superconducting qubit to optical photon, which is an important step toward the quantum state transfer between distant superconducting qubits interconnected over optical fibers.
\\
\\
\end{abstract}

\maketitle

\section*{Introduction}
\begin{figure*}[!t]
\centering
\includegraphics[width=2\columnwidth]{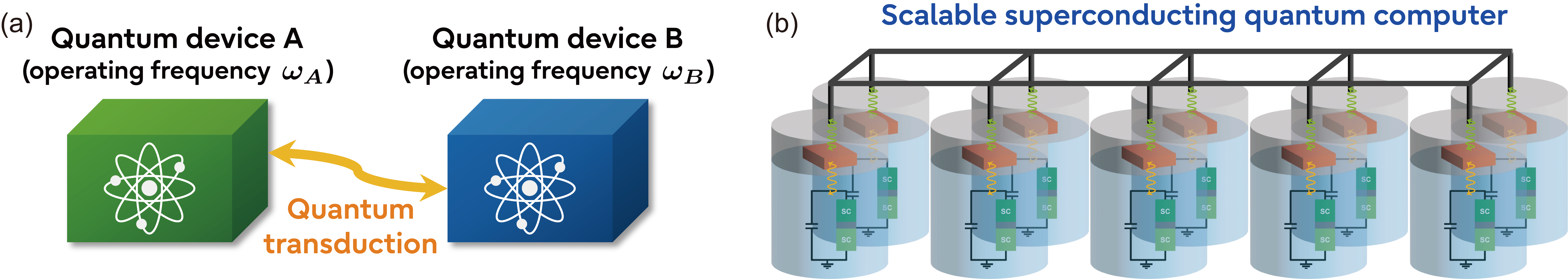}
\caption{{\bf Quantum interconnects via the quantum transduction.}
(a) Schematic illustration of a quantum interconnect. In general, when the operating frequencies of the two quantum devices $\omega_A$ and $\omega_B$ are different, a quantum transducer (or equivalently quantum frequency converter) is required in order to interconnect the two quantum devices.
(b) Schematic illustration of a scalable superconducting quantum computer system with the dilution refrigerators interconnected over optical fibers via the microwave-to-optical quantum transduction.
}\label{Fig-Quantum-Interconnects}
\end{figure*}

The quantum transduction, or equivalently quantum frequency conversion, which enables the interconnects between quantum devices such as quantum processors and quantum memories, is a vital technology for the realization of, e.g., quantum networks, distributed quantum computing, and quantum repeaters \cite{Wehner2018,Awschalom2021,Azuma2023,Beukers2024} (see Fig.~\ref{Fig-Quantum-Interconnects}).
When the frequency difference between two quantum devices is within the same order of magnitude in the microwave domain around $1\textrm{--}10\, \mathrm{GHz}$, a quantum transducer with a transduction efficiency higher than $99 \%$ and a low noise in the quantum regime has been developed using the Josephson parametric converter \cite{Bergeal2010,Abdo2011,Abdo2013,Abdo2014,Andrews2015} for superconducting quantum circuits.
On the other hand, the quantum transduction in the optical domain such as between telecom ($\approx 200\, \mathrm{THz}$) and visible ($\approx 400\, \mathrm{THz}$) lights is challenging.
While recent experiments have demonstrated the transduction efficiencies of about $0.1\textrm{--}0.5$ via nonlinear optical interactions \cite{Rakher2010,Ikuta2011,Guo2016,Bruch2018,Wang2018,Lu2020,Murakami2023}, it is still not easy to obtain the transduction efficiency higher than $0.5$.
Achieving a high-efficiency low-noise quantum transduction becomes more difficult when the frequency difference between two quantum devices is as large as between microwave and optical frequency ranges, which is the focus of this review.
\begin{figure*}[!t]
\centering
\includegraphics[width=2\columnwidth]{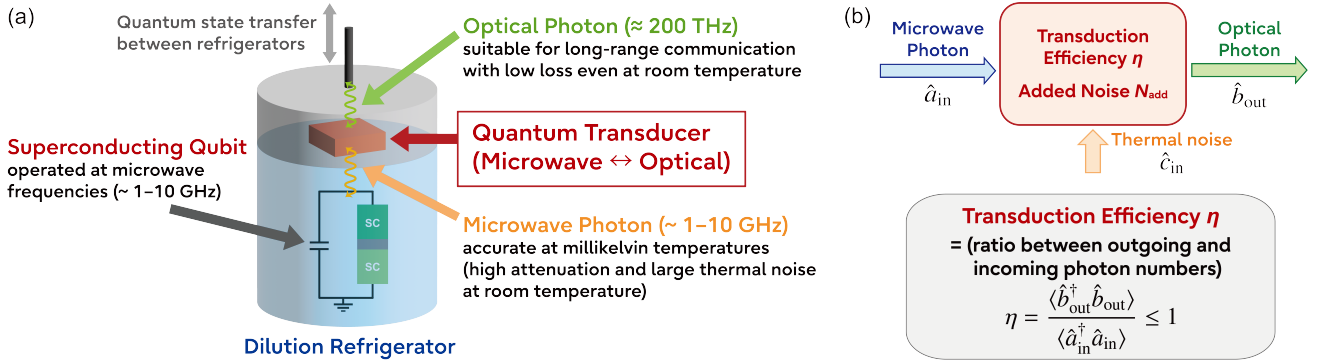}
\caption{{\bf The microwave-to-optical quantum transduction.}
(a) Schematic illustration of the microwave-to-optical quantum transduction in a superconducting quantum computing system.
The electric circuit in the dilution refrigerator represents a transmon qubit with two superconductors (SCs) colored green forming a Josephson junction.
Typically, the transducer is placed in the millikelvin stage of the refrigerator.
(b) A model of a direct, one-way quantum transduction characterized by the transduction efficiency $\eta$ and the added noise $N_{\mathrm{add}}$.
For clarity, the transduction from microwave photon to optical photon is described here.
The same argument is applied to the transduction from optical photon to microwave photon.
}\label{Fig-Quantum-transduction}
\end{figure*}

Superconducting qubits have been widely considered a promising constituent of a quantum processor to realize practical quantum computers \cite{Neill2018,Arute2019,Acharya2025}.
To date, several hundreds of superconducting qubits have been implemented in a single dilution refrigerator \cite{Krinner2019,Kjaergaard2020}.
The wiring technology becomes more complicated and the resultant thermal load becomes larger as the number of superconducting qubits grows, which indicates the physical limitation of the maximal number of qubits that can be placed in a dilution refrigerator.
On the other hand, roughly speaking, one logical qubit can be constructed with error-correcting codes from approximately $1000$ physical qubits \cite{Fowler2012}.
For practical calculations, approximately $10\textrm{--}1000$ logical qubits are required \cite{Bravyi2022,Mohseni2024}.
Therefore, one possible solution for this requirement with superconducting qubits is interconnecting dilution refrigerators.

For this reason, the microwave-to-optical quantum transduction is of particular interest in the field of superconducting quantum computing \cite{Lauk2020,Lambert2020,Han2021}, in pursuit of realizing large-scale quantum computers with fault-tolerance.
In Fig.~\ref{Fig-Quantum-transduction}(a), a schematic illustration of a quantum transducer system with superconducting qubits is presented.
In the current technology, optical fibers at the telecom frequency range are suitable for long-range communication with a low loss even at room temperature.
On the other hand, the superconducting qubits are operated by microwaves at low temperatures, typically in the millikelvin regime.
Also, it is difficult to transfer quantum states over long distances at microwave frequencies due to the high attenuation and large thermal noise at room temperature.
However, due to the large frequency difference between microwave and optical ranges, a direct frequency conversion of quantum signals between these two frequency ranges is almost impossible.
The quantum transduction is therefore done via the interactions with the intermediate bosonic mode(s) or via a nonlinear interaction process of photons.

The transduction efficiency $\eta$ can generically be defined by the ratio between incoming and outgoing photon numbers or photon fluxes [see Fig.~\ref{Fig-Quantum-transduction}(b)].
In the context of the quantum capacity (which we shall review in more detail), it has been suggested that a direct, one-way quantum transducer must satisfy a high transduction efficiency of $\eta>1/2$ as well as a low added noise in the quantum regime $N_{\mathrm{add}}\ll 1$ \cite{Holevo2001,Wolf2007,Weedbrook2012,Wang2022} in order to realize quantum state transfer.
Thus, achieving a quantum transduction with a high transduction efficiency and a low added noise is the necessary direction.
Here, it should be noted that reaching the quantum regime of $N_{\mathrm{add}} < 1$ itself is in principle a route to realize quantum communications, since arbitrarily low transduction efficiency of $\eta<1/2$ can be supplemented by the so-called heralded entanglement generation \cite{Duan2001,Muralidharan2016,Zhong2020,Krastanov2021} which has been considered in the field of quantum repeaters.
However, the experimental setups based on such a protocol may be more complicated than the direct transducers which are the focus of this review, because two-way classical communication signaling is required in the former.

In this review, we overview recent theoretical and experimental studies on the quantum transduction between microwave and optical photons.
The organization of this review is as follows.
Firstly, we briefly review the concept of the quantum capacity which fundamentally gives the lower bound of the transduction efficiency required for quantum state transfer.
Secondly, we describe a generic theory for the quantum transduction employing the input-output formalism, from which the essential quantities characterizing the transduction, i.e., the expressions for the transduction efficiency, the added noise, and the transduction bandwidth are derived.
Thirdly, we review the major transduction methods that have been experimentally demonstrated, focusing on the transduction via the optomechanical effect, the electro-optic effect, the magneto-optic effect, and the atomic ensembles.
Finally, we briefly review the recent experimental progress on the quantum transduction from superconducting qubit to optical photon, which is an
important step toward the quantum state transfer between superconducting qubits interconnected over optical fibers.

\section*{Quantum capacity}
In this section, we briefly consider how the threshold of the transduction efficiency $\eta=1/2$, above which a quantum state transfer between distant quantum devices is enabled, is derived.
The key quantity is the quantum capacity, which represents the maximal number of qubits that can be transmitted faithfully through a quantum channel \cite{Holevo2001,Wolf2007}.
A quantum transducer which we are focusing on in this review can be modeled as a Gaussian lossy bosonic channel \cite{Holevo2001,Wolf2007,Weedbrook2012}.
In a Gaussian lossy bosonic channel, the input and output photons at a given frequency $\omega$ are related by
\begin{align}
\hat{b}_{\mathrm{out}}(\omega)=\sqrt{\eta(\omega)}\hat{a}_{\mathrm{in}}(\omega)+\sqrt{1-\eta(\omega)}\hat{c}_{\mathrm{in}}(\omega),
\label{Eq-lossy-channel}
\end{align}
where $\eta$ is the transmissivity (corresponding to the transduction efficiency), $\hat{b}_{\mathrm{out}}$ ($\hat{a}_{\mathrm{in}}$) is the annihilation operator of the output (input) photons in a channel, and $\hat{c}_{\mathrm{in}}$ is the annihilation operator of the thermal (noisy) input photons of the environment with the average photon number $\bar{n}(\omega)=\int d\omega'/(2\pi)\, \langle\hat{c}_{\mathrm{in}}^\dag(\omega)\hat{c}_{\mathrm{in}}(\omega')\rangle$.
Note that, particularly in a pure-loss channel, we have $N_{b,\mathrm{out}}(\omega)=\eta(\omega) N_{a,\mathrm{in}}(\omega)$ in thermal equilibrium, because $\bar{n}(\omega)=0$ by definition \cite{Weedbrook2012}.
Here, $N_{b,\mathrm{out}}$ ($N_{a,\mathrm{in}}$) is the average number of output (input) photons in a channel, which is defined in a similar way as $\bar{n}(\omega)$ above. 
Therefore, we see that $\eta(\omega)$ in Eq.~(\ref{Eq-lossy-channel}) indeed represents the transduction efficiency between input and output photons.
Equation~(\ref{Eq-lossy-channel}) can be interpreted as a beam splitter that mixes the input signal and the thermal noise.

For a one-way quantum communication [see Fig.~\ref{Fig-Quantum-transduction}(b)] such as the quantum transduction from microwave photon to optical photon and vice versa satisfying $\bar{n}(\omega)=0$, which can be described as a pure-loss channel, the quantum capacity is given analytically by \cite{Holevo2001,Wolf2007,Weedbrook2012}
\begin{align}
q_1(\omega)&=\mathrm{max}\left\{\log_2\left(\frac{\eta(\omega)}{1-\eta(\omega)}\right),0\right\}
\nonumber\\
&
\left\{
\begin{aligned}
&\to \infty\ \ \ (\eta(\omega)\to 1),\\
&>0\ \ \ (1/2<\eta(\omega)<1),\\
&= 0\ \ \ (0\le\eta(\omega)\le 1/2)
\label{Eq-Quantum-capacity}
\end{aligned}
\right.
\end{align}
at a given frequency $\omega$.
From this equation we see that $\eta(\omega)=1/2$ along with $\bar{n}(\omega)=0$ is the threshold for realizing a quantum state transfer.
In other words, not only improving the transduction efficiency $\eta(\omega)$ but also reducing the thermal noise $\bar{n}(\omega)$ is essential for quantum transducers of practical use.

Recently, an extended version of the (discrete-time) quantum capacity [Eq.~(\ref{Eq-Quantum-capacity})], the continuous-time quantum capacity, was proposed to characterize quantum transducers.
It is defined by \cite{Wang2022}
\begin{align}
Q_1=\int \frac{d\omega}{2\pi}\, q_1(\omega),
\end{align}
where the integration is done over the transduction bandwidth.
This quantity describes the maximal amount of quantum information that can be reliably transmitted through the transducer per unit time, while the original quantum capacity [Eq.~(\ref{Eq-Quantum-capacity})] quantifies the maximal amount of quantum information that can be reliably transmitted in a one-way channel.

\begin{figure*}[!t]
\centering
\includegraphics[width=2\columnwidth]{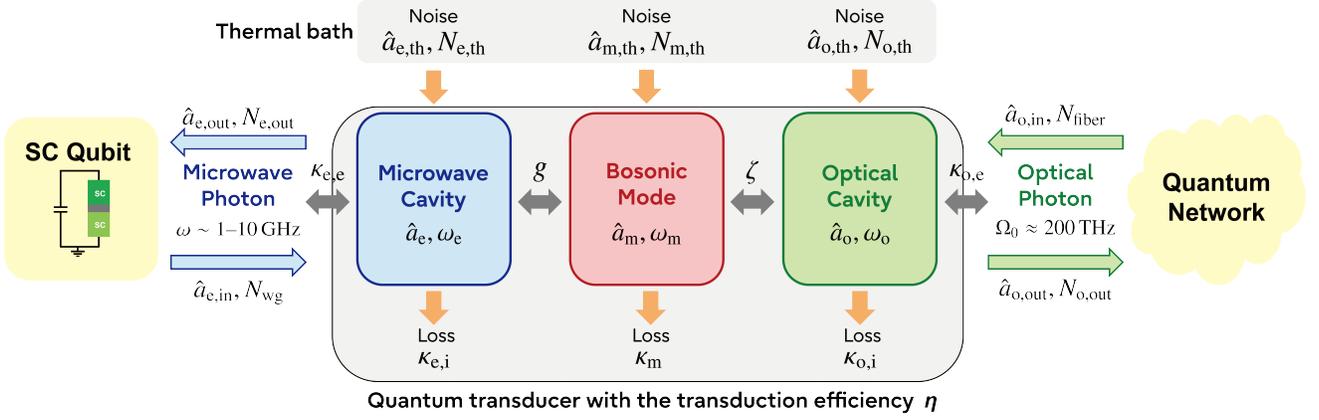}
\caption{{\bf Physical quantities and operators characterizing the one-stage quantum transduction.}
The quantum transduction mediated by one intermediate bosonic mode (such as phonon and magnon) can be described in terms of the operators ($\hat{a}_{\mathrm{e,in}}$, $\hat{a}_{\mathrm{e,out}}$, $\hat{a}_{\mathrm{m}}$, $\hat{a}_{\mathrm{o,in}}$, $\hat{a}_{\mathrm{o,out}}$, $\hat{a}_{\mathrm{e,th}}$, $\hat{a}_{\mathrm{m,th}}$, and $\hat{a}_{\mathrm{o,th}}$), frequencies ($\omega$, $\omega_{\mathrm{e}}$, $\omega_{\mathrm{m}}$, $\omega_{\mathrm{o}}$, and $\Omega_0$), coupling strengths ($\kappa_{\mathrm{e,e}}$, $g$, $\zeta$, and $\kappa_{\mathrm{o,e}}$), intrinsic loss rates ($\kappa_{\mathrm{e,i}}$, $\kappa_{\mathrm{m}}$, and $\kappa_{\mathrm{o,i}}$), and number of noise photons ($N_{\mathrm{wg}}$, $N_{\mathrm{e,out}}$, $N_{\mathrm{e,th}}$, $N_{\mathrm{m,th}}$, $N_{\mathrm{o,th}}$, $N_{\mathrm{fiber}}$, and $N_{\mathrm{o,out}}$).
Note that the intrinsic loss rates are understood as the coupling strengths to the thermal baths, although in this figure the intrinsic losses and thermal baths are drawn separately for readability.
Note also that the number of noise photons $N_{\mathrm{fiber}}$ and $N_{\mathrm{o,th}}$, which are associated with the itinerant optical photons $\hat{a}_{\mathrm{o,in}}$ (typically in optical fibers) and the thermal bath photons $\hat{a}_{\mathrm{o,th}}$, respectively, can generally be neglected because of the high photon frequency of $\approx 200\, \mathrm{THz}$.
}\label{Fig-one-stage-transduction}
\end{figure*}
%

\section*{Theory of the microwave-to-optical quantum transduction \label{Sec-Quantum-Transduction}}
In this section, we overview the basics of the theory of the microwave-to-optical quantum transduction.
Generically, a quantum transduction process can be categorized as an $N$-stage transduction, where $N$ is zero, one, two..., and so on \cite{Han2021}.
Here, an integer $N$ denotes the number of the intermediate modes involved in the transduction.
A generic model for the $N$-stage transduction can be obtained by mapping the transducer system to an effective electric circuit model \cite{Wang2022a}.
As concrete and common methods, we focus on the zero-stage and one-stage transduction in what follows.
We derive expressions for the transduction efficiency, the added noise, and the transduction bandwidth, which characterizes the transducer performance \cite{Zeuthen2020}.

\subsection*{General consideration \label{Theory-General-Consideration}}
We consider a generic description to obtain an expression for the transduction efficiency.
For concreteness, we here focus on the one-stage quantum transduction, i.e., the quantum transduction mediated by one intermediate bosonic mode (such as phonon and magnon) as shown in Fig.~\ref{Fig-one-stage-transduction}.
The formalism in what follows can be immediately generalized to the $n$-stage quantum transduction, including the zero-stage transduction via the electro-optic effect.

As illustrated in Fig.~\ref{Fig-one-stage-transduction}, let us define the vectors $\vec{c}=[\hat{a}_{\mathrm{e}},\hat{a}_{\mathrm{m}},\hat{a}_{\mathrm{o}}]^T$ and $\vec{c}_{\mathrm{in}}=[\hat{a}_{\mathrm{e,in}}, \hat{a}_{\mathrm{e,th}}, \hat{a}_{\mathrm{m,th}}, \hat{a}_{\mathrm{o,in}}, \hat{a}_{\mathrm{o,th}}]^T$.
Here, $\hat{a}_{\mathrm{e}}$, $\hat{a}_{\mathrm{m}}$, and $\hat{a}_{\mathrm{o}}$ are annihilation operators for the microwave cavity, intermediate bosonic, and optical cavity modes, respectively.
$\hat{a}_{\mathrm{e,in}}$ and $\hat{a}_{\mathrm{o,in}}$ ($\hat{a}_{\mathrm{e,out}}$ and $\hat{a}_{\mathrm{o,out}}$) are the input (output) itinerant microwave and optical photon operators, respectively.
$\hat{a}_{\mathrm{e,th}}$, $\hat{a}_{\mathrm{m,th}}$, and $\hat{a}_{\mathrm{o,th}}$ are the input thermal bath operators for the microwave cavity, intermediate bosonic, and optical cavity modes, respectively.
The equations of motion for the three modes ($\hat{a}_{\mathrm{e}}$, $\hat{a}_{\mathrm{m}}$, and $\hat{a}_{\mathrm{o}}$) are written as $\dot{\hat{c}}_j=(i/\hbar)[\hat{H}_{\mathrm{total}},\hat{c}_j]$, where $\hat{c}_j$ is the $j$-th component of $\vec{c}$ and $\hat{H}_{\mathrm{total}}$ is the total Hamiltonian of the system including the interaction Hamiltonian and the bath Hamiltonian.
See Supplementary Information for an explicit form of $\hat{H}_{\mathrm{total}}$ and for a derivation of the equations of motion.
Then, the equations of motion in the matrix representation can be written in a generic form 
\begin{align}
\dot{\vec{c}}=-A\vec{c}-B\vec{c}_{\mathrm{in}},
\label{equation-of-motion-generic-form}
\end{align}
where $A$ and $B$ are $3\times 3$ and $3\times 5$ matrices, respectively.
Employing the standard input-output formalism, the relation between the input and output photons can be written as
\begin{align}
\vec{c}_{\mathrm{out}}=\vec{c}_{\mathrm{in}}+B^T\vec{c},
\label{input-output-generic-form}
\end{align}
where $T$ is the transpose of a matrix and $\vec{c}_{\mathrm{out}}$ is defined from the output operators that are the counterparts of $\vec{c}_{\mathrm{in}}$.
Here,  the matrices $A$ and $B$ are given explicitly as
\begin{align}
A=
\begin{bmatrix}
i\omega_{\mathrm{e}}+\kappa_{\mathrm{e}}/2 & ig & 0\\
ig & i\omega_{\mathrm{m}}+\kappa_{\mathrm{m}}/2 & i\zeta\\
0 & i\zeta & -i\delta\omega_{\mathrm{o}}+\kappa_{\mathrm{o}}/2
\end{bmatrix}
\end{align}
and
\begin{align}
B=
\begin{bmatrix}
\sqrt{\kappa_{\mathrm{e,e}}} & \sqrt{\kappa_{\mathrm{e,i}}} & 0 & 0 & 0\\
0 & 0 & \sqrt{\kappa_{\mathrm{m}}} & 0 & 0\\
0 & 0 & 0 & \sqrt{\kappa_{\mathrm{o,e}}} & \sqrt{\kappa_{\mathrm{o,i}}}
\end{bmatrix},
\end{align}
respectively.
See Fig.~\ref{Fig-one-stage-transduction} for the definition of the variables introduced in the matrices $A$ and $B$.
Here, $-\delta \omega_{\mathrm{o}}=\omega_{\mathrm{o}}-\omega_{\mathrm{p}}$ is the detuning of the optical cavity frequency $\omega_{\mathrm{o}}$ from the pump frequency $\omega_{\mathrm{p}}$, and $\kappa_{\mathrm{e}}=\kappa_{\mathrm{e,e}}+\kappa_{\mathrm{e,i}}$ and $\kappa_{\mathrm{o}}=\kappa_{\mathrm{o,e}}+\kappa_{\mathrm{o,i}}$ are the total microwave and optical loss rates, respectively.
Combining Eqs.~(\ref{equation-of-motion-generic-form}) and (\ref{input-output-generic-form}), the scattering matrix $S$ that connects the input and output photons is introduced to obtain
\begin{align}
\vec{c}_{\mathrm{out}}(\omega)=S(\omega)\vec{c}_{\mathrm{in}}(\omega),
\label{Input-output-scattering-matrix}
\end{align}
where $S(\omega)=I_{5}-B^T[-i\omega I_{3}+A]^{-1}B$ with $I_{5}$ ($I_{3}$) being the $5\times 5$ ($3\times 3$) identity matrix and we have used the Fourier transform defined by $\vec{c}(t)=\int d\omega/(2\pi)\, e^{-i\omega t}\vec{c}(\omega)$.

Finally, introducing the susceptibilities $\chi_{\mathrm{e}}(\omega)=[-i(\omega-\omega_{\mathrm{e}})+\kappa_{\mathrm{e}}/2]^{-1}$, $\chi_{\mathrm{m}}(\omega)=[-i(\omega-\omega_{\mathrm{m}})+\kappa_{\mathrm{m}}/2]^{-1}$, and $\chi_{\mathrm{o}}(\omega)=[-i(\omega+\delta\omega_{\mathrm{o}})+\kappa_{\mathrm{o}}/2]^{-1}$, 
we can obtain physical quantities from the matrix elements $S_{i,j}$ such as the microwave-to-optical transduction efficiency $|S_{4,1}|^2=|S_{1,4}|^2$, the reflection coefficient for the itinerant microwave (optical) mode $|S_{1,1}|^2$ ($|S_{4,4}|^2$), and the added noise.

\subsection*{Transduction efficiency}
The transduction efficiency $\eta$ of the microwave-to-optical quantum transduction is defined from the matrix element of the scattering matrix [Eq.~(\ref{Input-output-scattering-matrix})] as the quantity representing the ratio between incoming (microwave) and outgoing (optical) photon numbers.
Namely, we have
\begin{align}
\eta(\omega)&=|S_{4,1}(\omega)|^2\nonumber\\
&=\left|\frac{g\zeta\sqrt{\kappa_{\mathrm{e,e}}}\sqrt{\kappa_{\mathrm{o,e}}}}{\chi_{\mathrm{e}}^{-1}(\omega)\chi_{\mathrm{o}}^{-1}(\omega)\chi_{\mathrm{m}}^{-1}(\omega)+g^2\chi_{\mathrm{o}}^{-1}(\omega)+\zeta^2\chi_{\mathrm{e}}^{-1}(\omega)}\right|^2.
\label{efficiency-definition}
\end{align}
Similarly, the transduction efficiency of the optical-to-microwave quantum transduction is given by $|S_{1,4}|^2$.
In our model, where the interaction Hamiltonian is of beam-splitter type, it turns out that $|S_{1,4}|^2=|S_{4,1}|^2=\eta$.
Under the resonance condition $\omega=-\delta\omega_{\mathrm{o}}=\omega_{\mathrm{e}}=\omega_{\mathrm{m}}$, we obtain an expression for the transduction efficiency in terms of the cooperativities,
\begin{align}
\eta=\eta_{\mathrm{e}}\eta_{\mathrm{o}}\frac{4C_{\mathrm{em}}C_{\mathrm{om}}}{(1+C_{\mathrm{em}}+C_{\mathrm{om}})^2},
\label{Transduction-efficiency-one-stage-transduction}
\end{align}
where $\eta_{\mathrm{o}}=\kappa_{\mathrm{o,e}}/\kappa_{\mathrm{o}}$, $\eta_{\mathrm{e}}=\kappa_{\mathrm{e,e}}/\kappa_{\mathrm{e}}$, and $C_{\mathrm{em}}=4g^2/(\kappa_{\mathrm{e}}\kappa_{\mathrm{m}})$ and $C_{\mathrm{om}}=4\zeta^2/(\kappa_{\mathrm{o}}\kappa_{\mathrm{m}})$ are, respectively, the cooperativity between microwave photons and the intermediate bosonic mode and the cooperativity between optical photons and the intermediate bosonic mode.
Another important quantity characterizing a transducer is the internal efficiency
\begin{align}
\eta_{\mathrm{in}}\equiv \frac{\eta}{\eta_{\mathrm{e}}\eta_{\mathrm{o}}},
\end{align}
which literally defines the efficiency that is independent of the external ports $\eta_{\mathrm{e}}$ and $\eta_{\mathrm{o}}$.
We see that, in order to achieve the unit efficiency $\eta=1$ in Eq.~(\ref{Transduction-efficiency-one-stage-transduction}), the condition such that $C_{\mathrm{em}}=C_{\mathrm{om}}\gg 1$, $\eta_{\mathrm{e}}\to 1$, and $\eta_{\mathrm{o}}\to 1$ is required.
Namely, high cooperativities and highly over-coupled external ports are required.

Next, let us consider the case of the zero-stage transduction via the electro-optic effect.
Under the resonance condition, we have an expression for the transduction efficiency in terms of the cooperativity (see Supplementary Information for a detailed derivation),
\begin{align}
\eta=\eta_{\mathrm{e}}\eta_{\mathrm{o}}\frac{4C_{\mathrm{eo}}}{(1+C_{\mathrm{eo}})^2},
\label{Transduction-efficiency-zero-stage-transduction}
\end{align}
where $C_{\mathrm{eo}}=4G_{\mathrm{eo}}^2/(\kappa_{\mathrm{e}}\kappa_{\mathrm{o}})$ is the cooperativity between microwave and optical photons.
Again, the internal efficiency is defined by $\eta_{\mathrm{in}}=\eta/\eta_{\mathrm{e}}\eta_{\mathrm{o}}$.
We see that, in order to achieve the unit efficiency $\eta=1$ in Eq.~(\ref{Transduction-efficiency-zero-stage-transduction}), the condition such that $C_{\mathrm{eo}}\to 1$, $\eta_{\mathrm{e}}\to 1$, and $\eta_{\mathrm{o}}\to 1$ is required.
In other words, unlike the one-stage transduction [Eq.~(\ref{Transduction-efficiency-one-stage-transduction})], a large cooperativity of $C_{\mathrm{eo}}\gg 1$ is not required.
Note that the expression for the transduction efficiency of the form of Eq.~(\ref{Transduction-efficiency-zero-stage-transduction}) can also be applied  to the cases of the transduction via an intermediate bosonic mode using only either a microwave or optical cavity.
In such cases, the cooperativity $C_{\mathrm{eo}}$ in Eq.~(\ref{Transduction-efficiency-zero-stage-transduction}) is replaced by the cooperativity between the cavity photons and the intermediate bosonic mode, $C_{\mathrm{em}}$ or $C_{\mathrm{om}}$.
Namely, we have
\begin{align}
\eta=\eta_{\mu}\eta_{\mathrm{m}}\frac{4C_{\mu\mathrm{m}}}{(1+C_{\mu\mathrm{m}})^2},
\label{Transduction-efficiency-zero-stage-transduction-2}
\end{align}
where $\mu=\mathrm{e}$ or $\mu=\mathrm{o}$ denotes the use of either microwave or optical cavity, $\eta_\mu=\kappa_{\mu,\mathrm{e}}/\kappa_{\mu}$, $\eta_{\mathrm{m}}=\tilde{G}/\kappa_{\mathrm{m}}$, and $C_{\mu\mathrm{m}}=4G_{\mu\mathrm{m}}^2/(\kappa_{\mu}\kappa_{\mathrm{m}})$.
Here, $\tilde{G}$ is the coupling strength between the intermediate bosonic mode and the itinerant photons whose cavity mode is absent, and $G_{\mu\mathrm{m}}$ is the coupling strength between the intermediate bosonic mode and the cavity photons.

\subsection*{Added noise}
Next, we take into account the presence of thermal noises of various origin which are illustrated in Fig.~\ref{Fig-one-stage-transduction}.
To this end, in the following we assume the input thermal noise correlators of the $\delta$-function form \cite{Aspelmeyer2014,Clerk2010},
\begin{align}
\langle\hat{c}_{\mu,\mathrm{th}}^\dag(t)\hat{c}_{\mu,\mathrm{th}}(t')\rangle=N_{\mu,\mathrm{th}}\, \delta(t-t'),
\end{align}
or equivalently, by performing the Fourier transform,
\begin{align}
\langle\hat{c}_{\mu,\mathrm{th}}^\dag(\omega)\hat{c}_{\mu,\mathrm{th}}(\omega')\rangle=2\pi N_{\mu,\mathrm{th}}\, \delta(\omega-\omega'),
\end{align}
where $\mu=\mathrm{e, m, o}$ and $N_{\mathrm{\mu,th}}=(e^{\hbar\omega_{\mu}/k_{\mathrm{B}}T_\mu}-1)^{-1}$ is the Bose distribution function at temperature $T_\mu$.
Here, note that the input operators $\hat{c}_{\mathrm{in}}(t)$ have the dimension of $\mathrm{T}^{-1/2}$ as can be seen from Eq.~(\ref{equation-of-motion-generic-form}).
The scattering matrix notation of Eq.~(\ref{Input-output-scattering-matrix}) can be written as
\begin{align}
\hat{a}_{\mathrm{o,out}}(\omega)&=\sqrt{\eta(\omega)}\hat{a}_{\mathrm{e,in}}(\omega)+\hat{d}_{\mathrm{in}}(\omega),\nonumber\\
\hat{a}_{\mathrm{e,out}}(\omega)&=\sqrt{\eta(\omega)}\hat{a}_{\mathrm{o,in}}(\omega)+\hat{e}_{\mathrm{in}}(\omega),
\label{Input-output-with-noise}
\end{align}
where we have introduced the thermal noise input operators $\hat{d}_{\mathrm{in}}$ and $\hat{e}_{\mathrm{in}}$.
Here, note that $\hat{d}_{\mathrm{in}}$ and $\hat{e}_{\mathrm{in}}$ in Eq.~(\ref{Input-output-with-noise}) correspond to $\sqrt{1-\eta}\hat{c}_{\mathrm{in}}$ in Eq.~(\ref{Eq-lossy-channel}).
Comparing Eqs.~(\ref{Input-output-scattering-matrix}) and (\ref{Input-output-with-noise}), we have
\begin{align}
\hat{d}_{\mathrm{in}}&=S_{4,1}\delta\hat{a}_{\mathrm{e,in}}+S_{4,2}\hat{a}_{\mathrm{e,th}}+S_{4,3}\hat{a}_{\mathrm{m,th}}+S_{4,4}\hat{a}_{\mathrm{o,in}}+S_{4,5}\hat{a}_{\mathrm{o,th}},\nonumber\\
\hat{e}_{\mathrm{in}}&=S_{1,1}\delta\hat{a}_{\mathrm{e,in}}+S_{1,2}\hat{a}_{\mathrm{e,th}}+S_{1,3}\hat{a}_{\mathrm{m,th}}+S_{1,5}\hat{a}_{\mathrm{o,th}},
\end{align}
where we have introduced an operator $\delta\hat{a}_{\mathrm{e,in}}$ denoting the thermal noise associated with the itinerant microwave photon $\hat{a}_{\mathrm{e,in}}$, i.e., the thermal noise present such as in the waveguide.
Since we can safely ignore the thermal noise of the optical cavity, i.e., $N_{\mathrm{o,th}}\approx 0$, as well as the thermal noise of the optical fiber, i.e., $N_{\mathrm{fiber}}\approx 0$ even at room temperature due to the high photon frequency $\omega_{\mathrm{o}}\approx 200\, \mathrm{THz}$.
The average numbers of the input thermal noise photons are then given by
\begin{align}
N_{\mathrm{o,out}}(\omega)&=\int \frac{d\omega'}{2\pi}\,\langle \hat{d}_{\mathrm{in}}^\dag(\omega) \hat{d}_{\mathrm{in}}(\omega')\rangle\nonumber\\
&=|S_{4,1}|^2N_{\mathrm{wg}}+|S_{4,2}|^2N_{\mathrm{e,th}}+|S_{4,3}|^2N_{\mathrm{m,th}},
\end{align}
\begin{align}
N_{\mathrm{e,out}}(\omega)&=\int \frac{d\omega'}{2\pi}\,\langle \hat{e}_{\mathrm{in}}^\dag(\omega) \hat{e}_{\mathrm{in}}(\omega')\rangle\nonumber\\
&=|S_{1,1}|^2N_{\mathrm{wg}}+|S_{1,2}|^2N_{\mathrm{e,th}}+|S_{1,3}|^2N_{\mathrm{m,th}},
\end{align}
where $N_{\mathrm{wg}}=(e^{\hbar\omega/k_{\mathrm{B}}T_{\mathrm{wg}}}-1)^{-1}$ is the Bose distribution function at temperature $T_{\mathrm{wg}}$ of the waveguide.
Here, note that these $N_{\mathrm{o,out}}$ and $N_{\mathrm{e,out}}$ are dimensionless quantities.
Precisely speaking, they are the spectral density that represents the number of thermal photons passing a given point per unit time per unit bandwidth \cite{Clerk2010}.
Thus, they are usually given in units of $\mathrm{s^{-1}}\, \mathrm{Hz^{-1}}$.
Under the resonance condition $\omega=-\delta\omega_{\mathrm{o}}=\omega_{\mathrm{e}}=\omega_{\mathrm{m}}$, the matrix elements are given explicitly as $|S_{4,1}|^2=\eta=\eta_{\mathrm{o}}\eta_{\mathrm{e}}\frac{4C_{\mathrm{om}}C_{\mathrm{em}}}{(1+C_{\mathrm{om}}+C_{\mathrm{em}})^2}$, $|S_{4,2}|^2=\eta_{\mathrm{o}}(1-\eta_{\mathrm{e}})\frac{4C_{\mathrm{om}}C_{\mathrm{em}}}{(1+C_{\mathrm{om}}+C_{\mathrm{em}})^2}$, $|S_{4,3}|^2=\eta_{\mathrm{o}}\frac{4C_{\mathrm{om}}}{(1+C_{\mathrm{om}}+C_{\mathrm{em}})^2}$, $|S_{1,1}|^2=\big|1-2\eta_{\mathrm{e}}\frac{1+C_{\mathrm{om}}}{1+C_{\mathrm{om}}+C_{\mathrm{em}}}\big|^2$, $|S_{1,2}|^2=\eta_{\mathrm{e}}(1-\eta_{\mathrm{e}})\frac{4(1+C_{\mathrm{om}})^2}{(1+C_{\mathrm{om}}+C_{\mathrm{em}})^2}$, and $|S_{1,3}|^2=\eta_{\mathrm{e}}\frac{4C_{\mathrm{em}}}{(1+C_{\mathrm{om}}+C_{\mathrm{em}})^2}$.

The added noise $N_{\mathrm{add}}$ is defined as the average number of photons added to the average number of the input signal photons, i.e.,  $N_{\mathrm{e,in}}(\omega)=\int d\omega'/(2\pi)\, \langle\hat{a}^\dag_{\mathrm{e,in}}(\omega)\hat{a}_{\mathrm{e,in}}(\omega')\rangle$ or $N_{\mathrm{o,in}}(\omega)=\int d\omega'/(2\pi)\, \langle\hat{a}^\dag_{\mathrm{o,in}}(\omega)\hat{a}_{\mathrm{o,in}}(\omega')\rangle$ in Eq.~(\ref{Input-output-with-noise}), which is thus given by $N_{\mathrm{add,\, M\to O}}(\omega)\equiv N_{\mathrm{o,out}}(\omega)/\eta(\omega)$ for the microwave-to-optical transduction and $N_{\mathrm{add,\, O\to M}}(\omega)\equiv N_{\mathrm{e,out}}(\omega)/\eta(\omega)$ for the optical-to-microwave transduction.
In this sense, these quantities are also called the ``input-referred'' added noise.
Explicitly, we have in the on-resonance condition
\begin{align}
N_{\mathrm{add,\, M\to O}}=N_{\mathrm{wg}}+\left(\frac{1}{\eta_{\mathrm{e}}}-1\right)N_{\mathrm{e,th}}+\frac{1}{\eta_{\mathrm{e}}C_{\mathrm{em}}}N_{\mathrm{m,th}},
\label{Added-noise-optical-out}
\end{align}
\begin{align}
N_{\mathrm{add,\, O\to M}}=&\, \frac{1}{\eta_{\mathrm{o}}\eta_{\mathrm{e}}}\frac{\left|C_{\mathrm{em}}+(1-2\eta_{\mathrm{e}})(1+C_{\mathrm{om}})\right|^2}{4C_{\mathrm{om}}C_{\mathrm{em}}}N_{\mathrm{wg}}\nonumber\\
&+\frac{1-\eta_{\mathrm{e}}}{\eta_{\mathrm{o}}}\frac{(1+C_{\mathrm{om}})^2}{C_{\mathrm{om}}C_{\mathrm{em}}}N_{\mathrm{e,th}}+\frac{1}{\eta_{\mathrm{o}}C_{\mathrm{om}}}N_{\mathrm{m,th}}.
\label{Added-noise-microwave-out}
\end{align}
We see from Eqs.~(\ref{Added-noise-optical-out}) and (\ref{Added-noise-microwave-out}) that the contribution from the microwave thermal noise $N_{\mathrm{e,th}}$ can be reduced by realizing a highly over-coupled microwave port $\eta_{\mathrm{e}}\to 1$.
As for reduction of the intermediate-mode thermal noise $N_{\mathrm{m,th}}$, high cooperativities $C_{\mathrm{em}}\gg 1$ and $C_{\mathrm{om}}\gg 1$ are required.

In the case of the zero-stage transduction, the added noises, $N_{\mathrm{add,\, M\to O}}$ for the microwave-to-optical transduction and $N_{\mathrm{add,\, O\to M}}$ for the optical-to-microwave transduction, are obtained as
\begin{align}
N_{\mathrm{add,\, M\to O}}=N_{\mathrm{wg}}+\left(\frac{1}{\eta_{\mathrm{e}}}-1\right)N_{\mathrm{e,th}},
\label{Added-noise-optical-out-zerostage}
\end{align}
\begin{align}
N_{\mathrm{add,\, O\to M}}=\frac{1}{\eta_{\mathrm{e}}\eta_{\mathrm{o}}}\frac{\left|1-2\eta_{\mathrm{e}}+C_{\mathrm{eo}}\right|^2}{4C_{\mathrm{eo}}}N_{\mathrm{wg}}+\frac{1-\eta_{\mathrm{e}}}{\eta_{\mathrm{o}}}\frac{1}{C_{\mathrm{eo}}}N_{\mathrm{e,th}}.
\label{Added-noise-microwave-out-zerostage}
\end{align}
We see from Eqs.~(\ref{Added-noise-optical-out-zerostage}) and (\ref{Added-noise-microwave-out-zerostage}) that  the contribution from the microwave thermal noise $N_{\mathrm{e,th}}$ can be reduced by a highly over-coupled microwave port $\eta_{\mathrm{e}}\to 1$.
See Supplementary Information for a derivation of the added noise in the zero-stage quantum transduction.

\begin{figure}[!t]
\centering
\includegraphics[width=\columnwidth]{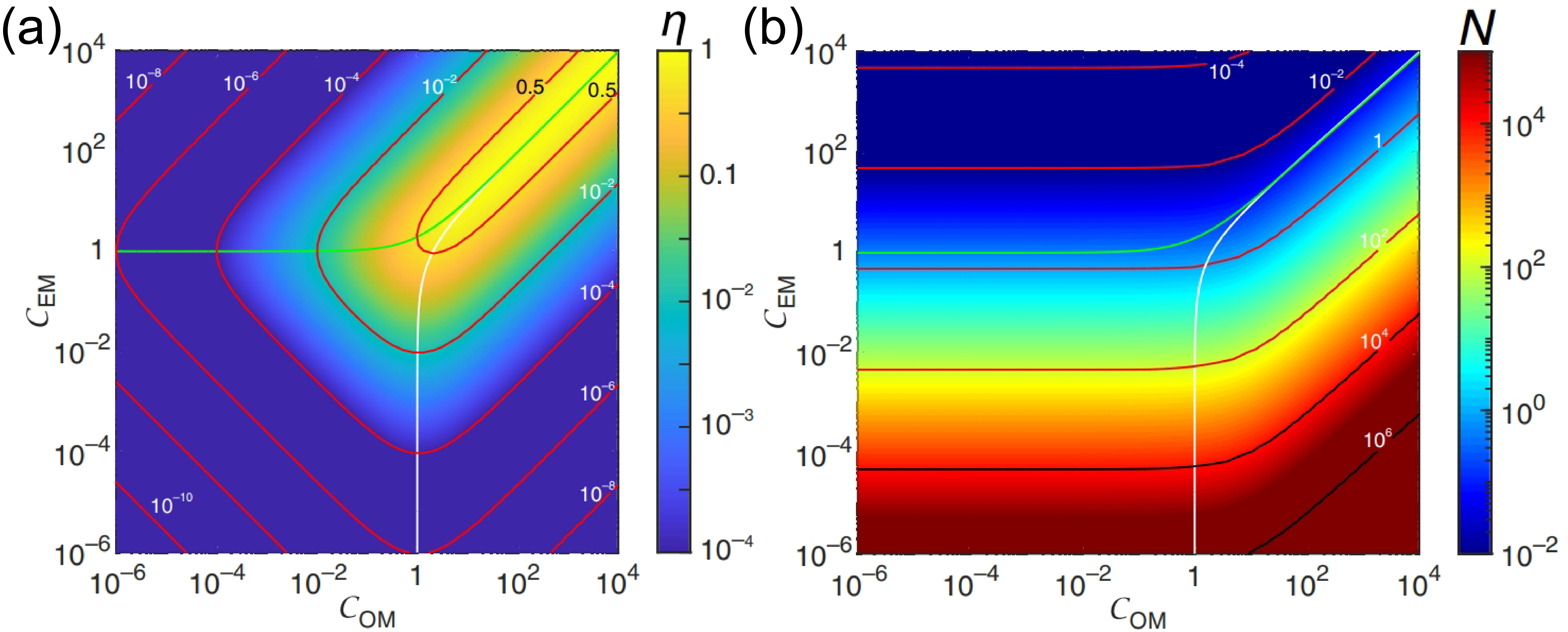}
\caption{{\bf Trade-off between high efficiency and low added noise.}
(a) Transduction efficiency $\eta$ and (b) added noise $N_{\mathrm{add}}$ as a function of the cooperativities $C_{\mathrm{om}}$ and $C_{\mathrm{em}}$ of the microwave-to-optical transduction in a piezo-optomechanical system.
Adapted with permission from Ref.~\onlinecite{Wu2020}.
Copyright 2020 American Physical Society.
}\label{Fig-Trade-off}
\end{figure}
\subsection*{Trade-off between high efficiency and low added noise}
It is essential to achieve a high transduction efficiency of $\eta>1/2$ and a low added noise of $N_{\mathrm{add}}\ll 1$ simultaneously for the quantum state transfer and remote entanglement between distant (superconducting) qubits.
Here, let us see that there is a trade-off between high efficiency and low added noise \cite{Zeuthen2020,Wu2020}.
Using the commutation relations of the itinerant fields such as $[\hat{a}_{\mathrm{o,out}}(\omega), \hat{a}_{\mathrm{o,out}}^\dag(\omega')]=2\pi\delta(\omega-\omega')$ \cite{Caves1982}, we obtain from Eq.~(\ref{Input-output-with-noise}) a constraint for the microwave-to-optical transduction
\begin{align}
1=\eta(\omega)+|S_{4,2}(\omega)|^2+|S_{4,3}(\omega)|^2+|S_{4,4}(\omega)|^2+|S_{4,5}(\omega)|^2,
\end{align}
and a constraint for the optical-to-microwave transduction
\begin{align}
1=\eta(\omega)+|S_{1,1}(\omega)|^2+|S_{1,2}(\omega)|^2+|S_{1,3}(\omega)|^2+|S_{1,5}(\omega)|^2.
\end{align}
These constraints relate the transduction efficiency to the added noise in a transducer system, and from which we have naturally that $\eta\le 1$.
Note here that we have restricted ourselves to the fully resolved-sideband regime ($4\omega_{\mathrm{m}}\gg \kappa_{\mathrm{e}}, \kappa_{\mathrm{o}}$) where the interaction Hamiltonian takes the form of a beam-splitter type interaction.
Outside the resolved sideband regime, there can be amplification noise giving rise to $\eta>1$ \cite{Zeuthen2020}due to the two-mode squeezing interaction of the form $H_{\mathrm{int}}=\hbar g(\hat{a}_{\mathrm{e}}^\dag\hat{a}_{\mathrm{m}}^\dag+\mathrm{H.c.})$ and/or $H_{\mathrm{int}}=\hbar \zeta(\hat{a}_{\mathrm{o}}^\dag\hat{a}_{\mathrm{m}}^\dag+\mathrm{H.c.})$.
In addition, such amplification noise arising from a finite sideband resolution gives a lower bound of the added noise $N_{\mathrm{add}}$ \cite{Zeuthen2020,Wu2020}.

Figure~\ref{Fig-Trade-off} shows the transduction efficiency $\eta$ and the added noise $N_{\mathrm{add}}$ as a function of the cooperativities $C_{\mathrm{om}}$ and $C_{\mathrm{em}}$ of the microwave-to-optical transduction in a piezo-optomechanical system  \cite{Wu2020}.
In this analysis, a finite sideband resolution is taken into account and the resulting optical amplification noise (Raman scattering noise) gives a lower bound on $N_{\mathrm{add}}$.
From this figure, we see that in practice the unit transduction efficiency $\eta=1$ and the zero added noise $N_{\mathrm{add}}=0$ cannot be realized simultaneously.
Note, however, that a transduction efficiency of $\eta>1/2$ and an added noise in the quantum-enabled regime of $N_{\mathrm{add}}< 1$ can be achieved simultaneously.

\subsection*{Transduction bandwidth}
Now, we discuss the transduction bandwidth, which determines the capacity of a given quantum communication channel.
A higher transduction bandwidth is vital for practical use, because the higher the transduction bandwidth becomes, the more the number of quantum communication channels that can be used at the same time becomes.
The transduction bandwidth is determined by the decay rate of the slowest mode in the transduction chain.
Therefore, for the zero-stage quantum transduction such as the one with the electro-optic effect, the problem is basically simple in the low cooperativity regime: the transduction bandwidth is given by $\mathrm{min}[\kappa_{\mathrm{e}}, \kappa_{\mathrm{o}}]$ \cite{Han2021}.
On the other hand, for a one-stage quantum transduction, the linewidth of the intermediate bosonic mode is generically narrower than those of the microwave and optical cavity modes, $\kappa_{\mathrm{e}}$ and $\kappa_{\mathrm{o}}$.
For this reason, in the following, let us take a look at the dynamically broadened linewidth of the intermediate bosonic mode.

We start by rewriting a generic expression for the transduction efficiency $\eta$ [Eq.~(\ref{efficiency-definition})] by approximating the susceptibilities $\chi_{\mathrm{e}}(\omega)$ and $\chi_{\mathrm{o}}(\omega)$ nearly on resonance ($\omega=-\delta\omega_{\mathrm{o}}=\omega_{\mathrm{e}}=\omega_{\mathrm{m}}$) by the Lorentzian functions $\chi_{\mathrm{e}}(\omega)\approx \frac{\kappa_{\mathrm{e}}/2}{(\omega-\omega_{\mathrm{e}})^2+(\kappa_{\mathrm{e}}/2)^2}$ and $\chi_{\mathrm{o}}(\omega)\approx \frac{\kappa_{\mathrm{o}}/2}{(\omega+\delta\omega_{\mathrm{o}})^2+(\kappa_{\mathrm{o}}/2)^2}$.
Introducing the decay rates $\Gamma_{\mathrm{em}}(\omega)=2g^2\chi_{\mathrm{e}}(\omega)$ and $\Gamma_{\mathrm{om}}(\omega)=2\zeta^2\chi_{\mathrm{o}}(\omega)$, the transduction efficiency $\eta$ is written as
\begin{align}
\eta(\omega)\approx\eta_{\mathrm{e}}\eta_{\mathrm{o}}\frac{\Gamma_{\mathrm{em}}\Gamma_{\mathrm{om}}}{(\omega-\omega_{\mathrm{m}})^2+[(\kappa_{\mathrm{m}}+\Gamma_{\mathrm{em}}+\Gamma_{\mathrm{om}})/2]^2}.
\label{Transduction-bandwidth-one-stage}
\end{align}
Here, we see that the quantity in the denominator,
\begin{align}
\Delta\omega=\kappa_{\mathrm{m}}+\Gamma_{\mathrm{em}}+\Gamma_{\mathrm{om}},
\end{align}
indeed represents the transducer bandwidth, i.e., the dynamically broadened linewidth of the intermediate bosonic mode.
Note that, under the resonance condition, we have $\Gamma_{\mathrm{em}}=\kappa_{\mathrm{m}}C_{\mathrm{em}}$ and $\Gamma_{\mathrm{om}}=\kappa_{\mathrm{m}}C_{\mathrm{om}}$, reproducing the expression for the transduction efficiency in Eq.~(\ref{Transduction-efficiency-one-stage-transduction}).

The above discussion can also be applied to the zero-stage transduction.
To see this, let us for concreteness consider the transduction via an intermediate bosonic mode with an optical cavity, which applies to the one utilizing the GHz piezo-optomechanical effect.
In this case, the linewidth of the bosonic mode is narrower than that of the optical cavity.
Then, the transduction efficiency is written as
\begin{align}
\eta(\omega)\approx\eta_{\mathrm{o}}\eta_{\mathrm{m}}\frac{\Gamma_{\mathrm{om}}}{(\omega-\omega_{\mathrm{m}})^2+[(\kappa_{\mathrm{m}}+\Gamma_{\mathrm{om}})/2]^2},
\label{Transduction-bandwidth-zero-stage}
\end{align}
from which the transduction bandwidth is given by $\Delta\omega=\kappa_{\mathrm{m}}+\Gamma_{\mathrm{om}}$ with $\Gamma_{\mathrm{om}}=\kappa_{\mathrm{m}}C_{\mathrm{om}}$.
Again, on resonance we reproduce the expression for the transduction efficiency in Eq.~(\ref{Transduction-efficiency-zero-stage-transduction-2}).
Note that the form of Eq.~(\ref{Transduction-bandwidth-zero-stage}) can also describe the transduction via the electro-optic effect, giving rise to the bandwidth $\Delta\omega=\kappa_{\mathrm{e}}+\Gamma_{\mathrm{eo}}$, where we have assumed that $\kappa_{\mathrm{e}}<\kappa_{\mathrm{o}}$.

As an alternative approach, it is also convenient to make the analogy of electric circuits in order to characterize a quantum transducer.
For concreteness, we focus on the transduction via the optomechanical effect.
The decay rate (FWHM) of a piezoelectric circuit can be defined by $\kappa_{\mathrm{m}}=Z_{\mathrm{m}}/L_{\mathrm{m}}$ with $Z_{\mathrm{m}}$ the impedance and $L_{\mathrm{m}}$ the inductance of the circuit \cite{Zeuthen2018,Wu2020}.
The bandwidth of a quantum transducer utilizing the optomechanical effect is then given by the dynamically broadened linewidth of the intermediate phonon mode \cite{Wu2020}
\begin{align}
\Delta\omega&=(R_{\mathrm{m}}+R_{\mathrm{em}}+R_{\mathrm{om}})/L_{\mathrm{m}}\nonumber\\
&=\kappa_{\mathrm{m}}(1+C_{\mathrm{em}}+C_{\mathrm{om}}),
\label{Eq-Bandwidth}
\end{align}
where $R_{\mathrm{m}}$, $R_{\mathrm{em}}$, and $R_{\mathrm{om}}$ are the mechanical, electromechanical, and optomechanical impedance, respectively.
Here, the electromechanical and optomechanical cooperativities are defined by $C_{\mathrm{em}}=R_{\mathrm{em}}/R_{\mathrm{m}}$ and $C_{\mathrm{om}}=R_{\mathrm{om}}/R_{\mathrm{m}}$, respectively.
In the case of the transduction via the optomechanical effect, the rates of energy loss $\kappa_{\mathrm{m}}C_{\mathrm{em}}$ and $\kappa_{\mathrm{m}}C_{\mathrm{om}}$ are called the electromechanical and optomechanical decay rates, respectively.
From Eq.~(\ref{Eq-Bandwidth}) we see that the bandwidth is enhanced in systems with large electromechanical and optomechanical cooperativities $C_{\mathrm{em}}>1$ and $C_{\mathrm{om}}>1$, in addition to a larger intrinsic loss rate $\kappa_{\mathrm{m}}$ itself.

\begingroup
\renewcommand{\arraystretch}{1.3}
\begin{table*}[!t]
\caption{{\bf Schematic comparison of the microwave-to-optical quantum transduction via the MHz electro-optomechanical effect and via the GHz optomechanical effects (utilizing microdisk resonators and bulk acoustic resonators).}
NR, not reported.
--, not applicable.
SOI, silicon-on-insulator platform.
$^\clubsuit$ indicates a transduction from superconducting qubit to optical photon.
The efficiency $\eta$ denotes the total transduction efficiency [such as the one given by Eq.~(\ref{Transduction-efficiency-one-stage-transduction}) or Eq.~(\ref{Transduction-efficiency-zero-stage-transduction-2})] except for the transduction from superconducting qubit to optical photon.}
\begin{ruledtabular}
\begin{tabular}{cccccc||ccc}
Reference & Ref.~\onlinecite{Andrews2014} & Ref.~\onlinecite{Higginbotham2018} & Ref.~\onlinecite{Arnold2020} & Ref.~\onlinecite{Brubaker2022} & Ref.~\onlinecite{Delaney2022}$^\clubsuit$ & Ref.~\onlinecite{Han2020} & Ref.~\onlinecite{Yoon2023} & Ref.~\onlinecite{Blesin2024} \\
(Year) & (2014) & (2018) & (2020) & (2022) & (2022) & (2020) & (2023) & (2024)\\
\hline
System & Si$_3$N$_4$ & Si$_3$N$_4$ & SOI & Si$_3$N$_4$ & Si$_3$N$_4$ & AlN & CaF$_2$ & Si$_3$N$_4$\\
Frequency $\omega_{\mathrm{m}}$ & $560\, \mathrm{kHz}$ & $1.47\, \mathrm{MHz}$ & $11.8\, \mathrm{MHz}$ & $1.45\, \mathrm{MHz}$ & $1.45\, \mathrm{MHz}$ & $10.22\, \mathrm{GHz}$ & $11.37\, \mathrm{GHz}$ & $3.48\, \mathrm{GHz}$\\
Efficiency $\eta$ & $8.6\times 10^{-2}$ & 0.47 & $1.9\times 10^{-4}$ & 0.47 & $8\times 10^{-4}$ & $7.3\times 10^{-4}$ & $1.2\times 10^{-8}$ & $1.6\times 10^{-5}$\\
Cooperativity $C_{\mathrm{em}}$ & NR & 66 & 0.57 & 680 & $4.5\times 10^{3}$ & 7.4 & $5.6\times 10^{-8}$ & --\\
Cooperativity $C_{\mathrm{om}}$ & NR & 66 & 0.9 & 770 & $2.2\times 10^{4}$ & 0.4 & $\sim 1$ & NR\\
Added noise & NR & 38 & $\sim 100$ & 3.2 & 23 & NR & NR & NR\\
Bandwidth & $30\, \mathrm{kHz}$ & $12\, \mathrm{kHz}$ & $0.37\, \mathrm{kHz}$ & $2\, \mathrm{kHz}$ & $6.1\, \mathrm{kHz}$ & $1\, \mathrm{MHz}$ & $500\, \mathrm{kHz}$ & $25\, \mathrm{MHz}$\\
Temperature & $4\, \mathrm{K}$ & $35\, \mathrm{mK}$ & $50\, \mathrm{mK}$ & $40\, \mathrm{mK}$ & $40\, \mathrm{mK}$ & $0.9\, \mathrm{K}$ & $4\, \mathrm{K}$  & $300\, \mathrm{K}$\\
\end{tabular}
\end{ruledtabular}
\label{Table-MHz-Optomechanical}
\end{table*}
\endgroup
\begingroup
\renewcommand{\arraystretch}{1.3}
\begin{table*}[!t]
\caption{{\bf Schematic comparison of the microwave-to-optical quantum transduction via the GHz piezo-optomechanical effect in nanobeam geometries.}
NR, not reported.
--, not applicable.
SOI, silicon-on-insulator platform.
$^\clubsuit$ indicates a transduction from superconducting qubit to optical photon.
The efficiency $\eta$ denotes the total transduction efficiency [such as the one given by Eq.~(\ref{Transduction-efficiency-one-stage-transduction}) or Eq.~(\ref{Transduction-efficiency-zero-stage-transduction-2})] except for the transduction from superconducting qubit to optical photon.}
\begin{ruledtabular}
\begin{tabular}{ccccccccccc}
Reference & Ref.~\onlinecite{Vainsencher2016} & Ref.~\onlinecite{Forsch2020} & Ref.~\onlinecite{Jiang2020} & Ref.~\onlinecite{Mirhosseini2020}$^\clubsuit$ & Ref.~\onlinecite{Stockill2022} & Ref.~\onlinecite{Jiang2023} & Ref.~\onlinecite{Weaver2024} & Ref.~\onlinecite{vanThiel2025}$^\clubsuit$ & Ref.~\onlinecite{Zhao2025}\\
(Year) & (2016) & (2020) & (2020) & (2020) & (2022) & (2023) & (2024) & (2025) & (2025)\\
\hline
System & AlN & GaAs & LiNbO$_3$ & AlN & GaP & LiNbO$_3$ & LiNbO$_3$ & LiNbO$_3$ & SOI\\
Frequency $\omega_{\mathrm{m}}$ & $3.78\, \mathrm{GHz}$ & $2.7\, \mathrm{GHz}$ & $1.85\, \mathrm{GHz}$ & $5.16\, \mathrm{GHz}$ & $2.81\, \mathrm{GHz}$ & $3.60\, \mathrm{GHz}$ & $5.04\, \mathrm{GHz}$ & $5.19\, \mathrm{GHz}$ & $5.07\, \mathrm{GHz}$\\
Efficiency $\eta$ & $\sim 10^{-8}$ & $5.5\times 10^{-12}$ & $1.1\times 10^{-5}$ & $8.8\times 10^{-6}$ & $6.8\times 10^{-8}$ & $4.9\times 10^{-2}$ & $9\times 10^{-3}$ & $3.3\times 10^{-3}$ & $2.2\times 10^{-2}$\\
Cooperativity $C_{\mathrm{em}}$ & -- & -- & -- & -- & -- & 0.21 & 24.2 & NR & NR\\
Cooperativity $C_{\mathrm{om}}$ & $3\times 10^{-3}$ & 1.7 & $6.6\times 10^{-3}$ & $1.9\times 10^{-2}$ & 1.74 & 0.30 & NR & NR & NR\\
Added noise & NR & NR & NR & 0.57 & 0.55 & $\sim 100$ & 6.2 & $2\times 10^{3}$ & 0.94\\
Bandwidth & $\sim 1\, \mathrm{MHz}$ & $\sim 1\, \mathrm{MHz}$ & $\sim 1\, \mathrm{MHz}$ & $1\, \mathrm{MHz}$ & $\sim 0.1\, \mathrm{MHz}$ & $1.5\, \mathrm{MHz}$ & $14.8\, \mathrm{MHz}$ & $4.7\, \mathrm{MHz}$ & $88.9\, \mathrm{kHz}$\\
Temperature & $300\, \mathrm{K}$ & $20\, \mathrm{mK}$ & $300\, \mathrm{K}$ & $15\, \mathrm{mK}$ & $10\, \mathrm{mK}$ & $10\, \mathrm{mK}$ & $25\, \mathrm{mK}$ & $25\, \mathrm{mK}$ & $30\, \mathrm{mK}$
\end{tabular}
\end{ruledtabular}
\label{Table-GHz-Optomechanical}
\end{table*}
\endgroup
%

\section*{Current status of the microwave-to-optical quantum transduction}
In this section, we overview the recent experimental progress on the quantum transduction between microwave and optical photons.
We particularly focus on the four major transduction methods \cite{Lauk2020,Lambert2020,Han2021}; transduction via the optomechanical effect, the electro-optic effect, the magneto-optic effect, and the atomic ensembles .
Note that we do not distinguish between the ``microwave-to-optical transduction'' and the ``optical-to-microwave transduction'' unless otherwise mentioned, since most of the studies have explored the systems with the interaction Hamiltonian of beam-splitter type (in the sideband resolved regime) in which the transduction is bidirectional, i.e., the expressions for the transduction efficiency are identical.

\begin{figure}[!t]
\centering
\includegraphics[width=\columnwidth]{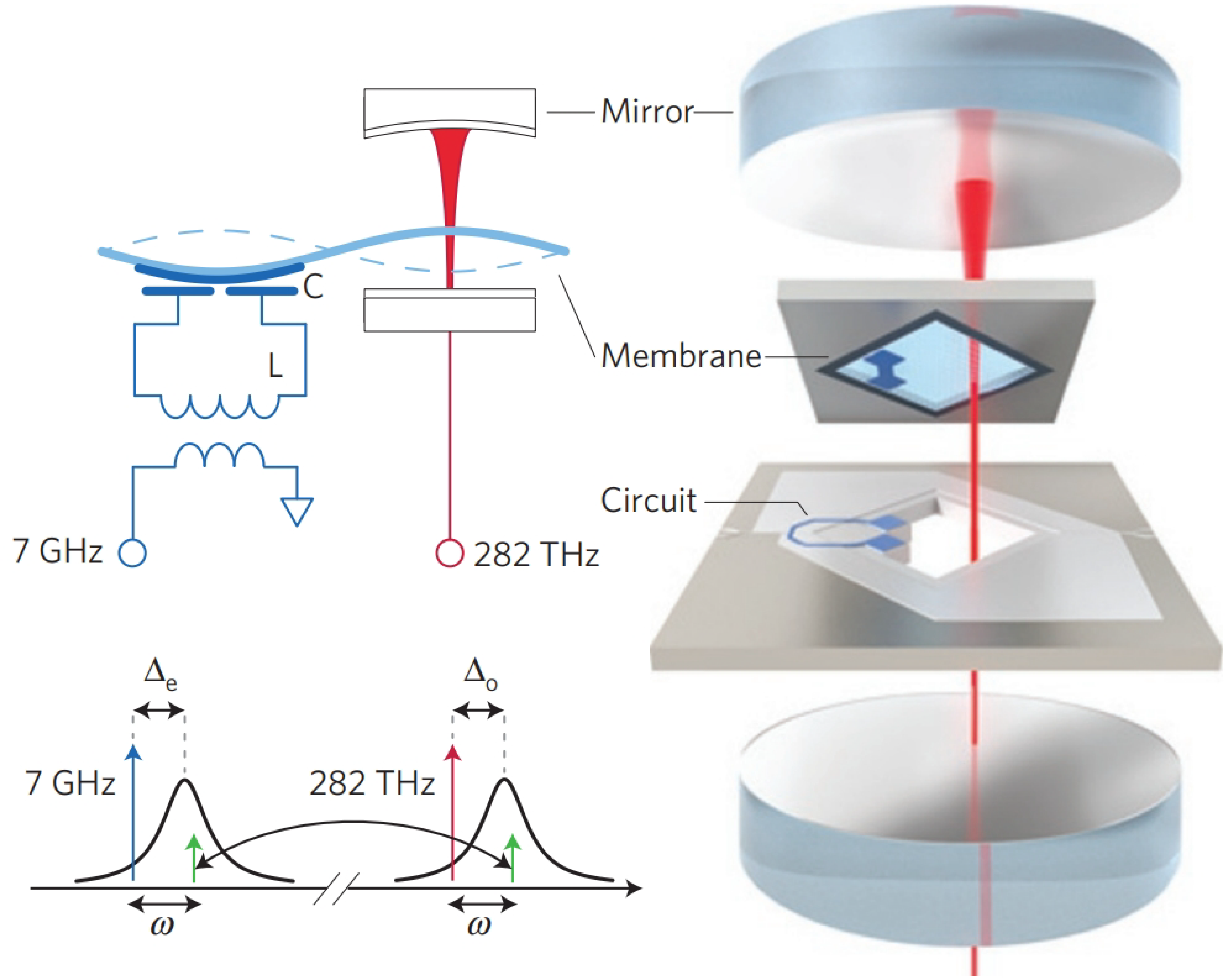}
\caption{{\bf MHz electro-optomechanical transducers.}
A typical membrane-based transducer device consists of a 3D optical cavity and a (superconducting) LC circuit that forms the microwave resonator.
Adapted with permission from Ref.~\onlinecite{Andrews2014}.
Copyright 2014 Springer Nature.
}\label{Fig-MHz-Optomechanical}
\end{figure}
\subsection*{Transduction via the optomechanical effect}
\subsubsection*{General argument}
The most widely investigated method for the microwave-to-optical quantum transduction is the one utilizing the optomechanical effect \cite{Aspelmeyer2014,Barzanjeh2022,Wang2024}, i.e., the coupling between optical photons and mechanical motion modes originating from the radiation pressure.
In other words, the phonons serve as the intermediate bosonic mode.
The quantum transduction via the optomechanical effect can be broadly classified into two methods.
One is with the electro-optomechanical effect, which is characterized by the use of the membranes with the resonance frequency in the MHz regime.
The other one is with the piezo-optomechanical effect, which is characterized by the use of the piezoelectric crystals with the resonance frequency in the GHz regime.
The MHz optomechanical transducers generically have the high transduction efficiency owing to high quality factors of the cavities and mechanical modes, whereas they can suffer from the high added noise and low transduction bandwidth due to the MHz mechanical resonance frequency.
The GHz optomechanical transducers in general have the low added noise owing to the GHz mechanical resonance frequency and can utilize the direct linear coupling between microwave photons and mechanical modes in piezoelectric crystals, whereas they can suffer from the fabrication and design complexity due to the nano-engineered structure of the beams.

To derive the coupling between photons and a mechanical vibrational mode, we begin with the total Hamiltonian of a  system with microwave and optical cavities, which is written as $H_0=\sum_{\mu=\mathrm{e,o}}\hbar\omega_{\mu}\hat{a}_{\mu}^\dag\hat{a}_\mu+\hbar\omega_{\mathrm{m}}\hat{a}_{\mathrm{m}}^\dag\hat{a}_{\mathrm{m}}$, where $\hat{a}_{\mathrm{e}}$ ($\hat{a}_{\mathrm{o}}$) and $\hat{a}_{\mathrm{m}}$ are the annihilation operators for the microwave (optical) cavity mode and the mechanical mode, respectively.
In general, the cavity resonance frequency $\omega_{\mu}$ can be modulated by a vibrational mode (mechanical motion) inside the cavity, such that $\omega_{\mu}(x)=\omega_{\mu,0}+x\partial \omega_{\mu}/\partial x+\cdots$.
Notice that the displacement $x$ can be quantized as $\hat{x}=x_{\mathrm{ZPF}}(\hat{a}_{\mathrm{m}}+\hat{a}_{\mathrm{m}}^\dag)$ \cite{Aspelmeyer2014}.
Here, $x_{\mathrm{ZPF}}$ is the mechanical zero-point fluctuation amplitude given by $x_{\mathrm{ZPF}}=\sqrt{\hbar/(2m_{\mathrm{eff}}\omega_{\mathrm{m}})}$ (with $m_{\mathrm{eff}}$ the effective mass of the mechanical oscillator).
Then, the interaction Hamiltonian is obtained as
\begin{align}
H_{\mathrm{OM}}=\hbar g_{\mathrm{em}}\hat{a}_{\mathrm{e}}^\dag\hat{a}_{\mathrm{e}}\left(\hat{a}_{\mathrm{m}}+\hat{a}_{\mathrm{m}}^\dag\right)+\hbar g_{\mathrm{om}}\hat{a}_{\mathrm{o}}^\dag\hat{a}_{\mathrm{o}}\left(\hat{a}_{\mathrm{m}}+\hat{a}_{\mathrm{m}}^\dag\right),
\label{Interaction-Hamiltonian-Optomechanical}
\end{align}
where the coupling strengths $g_{\mathrm{em}}$ and $g_{\mathrm{om}}$ are defined by $g_{\mathrm{em}}=g_{\mathrm{e}}x_{\mathrm{ZPF}}$ and $g_{\mathrm{om}}=g_{\mathrm{o}}x_{\mathrm{ZPF}}$, respectively.
Here, for a simple cavity of length $L_\mu$, we have $g_\mu=\omega_\mu/L_\mu$ \cite{Aspelmeyer2014}.
Note that this expression for the electromechanical coupling $g_{\mathrm{e}}$ is in general relevant for membrane systems and generally not applicable to other systems such as the one utilizing lumped element microwave resonators.
Now we rewrite the cavity photon operator $\hat{a}_\mu$ as $\hat{a}_\mu=\sqrt{n_\mu}+\delta\hat{a}_\mu$, where $n_\mu$ is the average intra-cavity pump photon number and $\delta\hat{a}_\mu$ is the fluctuating part.
Substituting this expression for $\hat{a}_\mu$ into Eq.~(\ref{Interaction-Hamiltonian-Optomechanical}), we have the linearized interaction Hamiltonian,
\begin{align}
H_{\mathrm{OM}}=&\ \hbar G_{\mathrm{em}}\left(\delta\hat{a}_{\mathrm{e}}+\delta\hat{a}_{\mathrm{e}}^\dag\right)\left(\hat{a}_{\mathrm{m}}+\hat{a}_{\mathrm{m}}^\dag\right)\nonumber\\
&+\hbar G_{\mathrm{om}}\left(\delta\hat{a}_{\mathrm{o}}+\delta\hat{a}_{\mathrm{o}}^\dag\right)\left(\hat{a}_{\mathrm{m}}+\hat{a}_{\mathrm{m}}^\dag\right),
\label{Interaction-Hamiltonian-Optomechanical-Linearized}
\end{align}
where $G_{\mathrm{em}}=g_{\mathrm{em}}\sqrt{n_{\mathrm{e}}}$ and $G_{\mathrm{om}}=g_{\mathrm{om}}\sqrt{n_{\mathrm{o}}}$.
Usually, $\delta\hat{a}_\mu$ is simply denoted as $\hat{a}_\mu$.
The Hamiltonian of the form of Eq.~(\ref{Interaction-Hamiltonian-Optomechanical-Linearized}) generically describes the coupling of the cavity radiation field (i.e., photons) to such a mechanical motion as membrane vibrations inside an optical cavity and localized modes in a piezoelectric crystal with a nanobeam structure \cite{Aspelmeyer2014}, as we review below.

\begin{figure*}[!t]
\centering
\includegraphics[width=2\columnwidth]{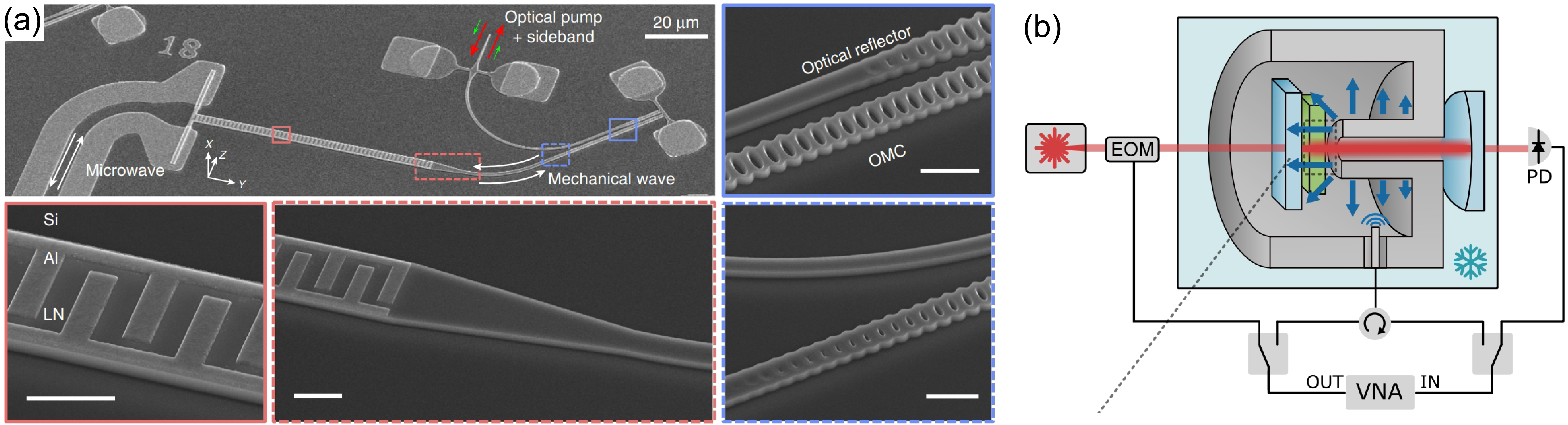}
\caption{{\bf GHz optomechanical transducers.}
(a) Scanning electron micrographs of a GHz piezo-optomechanical transducer device with a nanobeam structure.
Adapted from Ref.~\onlinecite{Jiang2020}.
Copyright 2020 The Authors, licensed under a Creative Commons Attribution (CC BY) license.
(b) Schematic illustration of a piezo-Brillouin optomechanical transducer device with a GHz bulk acoustic resonator.
Adapted with permission from Ref.~\onlinecite{Yoon2023}.
Copyright 2023 Optical Society of America.
}\label{Fig-GHz-Optomechanical}
\end{figure*}
\subsubsection*{MHz optomechanical transducers}
Table~\ref{Table-MHz-Optomechanical}I summarizes the experimental progress of the microwave-to-optical quantum transduction with the MHz optomechanical system \cite{Andrews2014,Higginbotham2018,Arnold2020,Brubaker2022,Delaney2022}.
Historically, the first experimental realization of the microwave-to-optical quantum transduction via the optomechanical effect was done with the MHz membrane system \cite{Andrews2014}.
The MHz membrane systems have so far been realized using Si$_3$N$_4$ \cite{Andrews2014,Higginbotham2018,Brubaker2022,Delaney2022} and SiN \cite{Bagci2014}.
The MHz optomechanical system has also been realized in a nanobeam geometry using silicon-on-insulator platform \cite{Arnold2020}.
As schematically shown in Fig.~\ref{Fig-MHz-Optomechanical}, a typical transducer device consists of a 3D optical cavity and a (superconducting) LC circuit that forms the microwave resonator.
In this case, the transduction efficiency $\eta$ is given by Eq.~(\ref{Transduction-efficiency-one-stage-transduction}).
One of the advantages of the membrane-based transducers is a high transduction efficiency resulting from large cooperativities.
The use of a 3D optical cavity such as the Fabry--P\'{e}rot cavity enables high quality factors of $Q_{\mathrm{o}}>10^8$.
Also, the membrane of Si$_3$N$_4$ can have extremely high mechanical quality factors of $Q_{\mathrm{m}}=\omega_{\mathrm{m}}/\kappa_{\mathrm{m}}>10^7$ \cite{Teufel2011,Yuan2015,Tsaturyan2017}.
So far, a transduction efficiency of $\eta=0.47$ with large cooperativities $C_{\mathrm{em}}\sim 100$ and $C_{\mathrm{om}}\sim 100$, which is close to the threshold $\eta=0.5$ required for quantum state transfer, has been recorded \cite{Higginbotham2018,Brubaker2022}.
Although the added noise coming from the low resonance frequency of $\omega_{\mathrm{m}}\sim 1\, \mathrm{MHz}$ of the membrane can be large in general [see Eqs.~(\ref{Added-noise-optical-out}) and (\ref{Added-noise-microwave-out})], a recent experiment has demonstrated a substantial reduction of the added noise to $3.2$ photons, keeping the transduction efficiency high ($\eta=0.47$) \cite{Brubaker2022}.
It is notable that a transduction from superconducting qubit to optical photon with $\eta=0.19$ (and the total quantum efficiency of $8\times 10^{-4}$) has been demonstrated \cite{Delaney2022}.

\subsubsection*{GHz optomechanical transducers}
Another promising method among the transduction via the optomechanical effect is to utilize nano-engineered crystals exhibiting the GHz phonon resonance frequency.
Owing to the resonance frequencies of such mechanical modes much higher than those of the MHz membrane systems, the thermal noise can be significantly reduced at millikelvin temperatures at which superconducting qubits are operated.
This direction of research was evolved from the discovery such that planar periodic nanobeam structures can be engineered to act simultaneously as a photonic and phononic crystal, resulting in a greatly enhanced optomechanical coupling strength \cite{Eichenfield2009}.
The optomechanical crystals can be fabricated with piezoelectric materials or non-piezoelectric materials.
The GHz optomechanical systems with the nanobeam structure have so far been realized using AlN \cite{OConnell2010,Bochmann2013,Vainsencher2016,Mirhosseini2020}, GaAs \cite{Balram2016,Forsch2020}, LiNbO$_3$ \cite{Jiang2020,Jiang2023,Weaver2024,vanThiel2025}, GaP \cite{Stockill2022,Hoenl2022}, and silicon-on-insulator platform \cite{Eichenfield2009,Zhao2025}.

One of the advantages of the GHz piezo-optomechanical systems is the linear coupling between phonons and microwave photons originating from the piezoelectric effect $H_{\mathrm{piezo}}=\int dV\, \Delta\bm{P}\cdot\bm{E}=\int dV\, [(\hat{\bm{e}}\cdot\bm{S})\cdot\bm{E}]$, which enables a direct coupling with superconducting quantum circuits.
Here, $\Delta\bm{P}$ is the strain-induced electric polarization, $\bm{E}$ is the applied electric field that is responsible for microwave photons, $\hat{\bm{e}}$ is the piezoelectric coefficient tensor, and $\bm{S}$ is the strain tensor.
The interaction Hamiltonian is written in terms of the phonon mode $\hat{a}_{\mathrm{m}}$ and the microwave photon mode $\hat{a}_{\mathrm{e}}$ as \cite{Han2016}
\begin{align}
H_{\mathrm{piezo}}=\hbar g_{\mathrm{piezo}}\left(\hat{a}_{\mathrm{e}}^\dag\hat{a}_{\mathrm{m}}+\hat{a}_{\mathrm{e}}\hat{a}_{\mathrm{m}}^\dag\right),
\label{Hamiltonian-piezo}
\end{align}
where the coupling strength $g_{\mathrm{piezo}}$ is given by $g_{\mathrm{piezo}}=\frac{e_{33}}{2\sqrt{\varepsilon_0 \rho}}\sqrt{\frac{\omega_{\mathrm{e}}}{\omega_{\mathrm{m}}}}\int_V\, dV \zeta_z(\bm{r})\frac{\partial}{\partial z}\chi_{z}(\bm{r})$.
Here, an electric field is applied along the $z$ direction, $e_{33}$ is the 33 component of the piezoelectric coefficient under the contracted Voigt notation, $\varepsilon_0$ is the vacuum permittivity, and $\rho$ is the mass density of the mechanical resonator, and $\zeta_z(\bm{r})$ [$\chi_{z}(\bm{r})$] is the $z$ component of the normalized electric (mechanical displacement) mode profile.

Table~\ref{Table-GHz-Optomechanical} summarizes the experimental progress of the microwave-to-optical quantum transduction with the GHz optomechanical system in nanobeam geometries  \cite{Vainsencher2016,Forsch2020,Jiang2020,Mirhosseini2020,Stockill2022,Jiang2023,Weaver2024,vanThiel2025,Zhao2025}.
Figure~\ref{Fig-GHz-Optomechanical}(a) shows scanning electron micrographs of a piezo-optomechanical transducer device \cite{Jiang2020}, in which the transduction process can be seen clearly.
A state-of-the-art platform is presented in, e.g., Ref.~\onlinecite{Weaver2024}, which features a transducer monolithically integrated with on-chip superconducting circuits.
The direct linear coupling between phonons and microwave photons [Eq.~(\ref{Hamiltonian-piezo})] indicates that it is both possible to use or not to use a microwave resonator (or equivalently, cavity).
Therefore, the GHz piezo-optomechanical systems can be divided into two categories: those with a microwave resonator and those without a microwave resonator.
In systems with a microwave resonator, the transduction efficiency is given by Eq.~(\ref{Transduction-efficiency-one-stage-transduction}).
On the other hand, in systems without a microwave resonator, the transduction efficiency is given by Eq.~(\ref{Transduction-efficiency-zero-stage-transduction-2}).
In such systems, the coupling between microwave photons and phonons is realized by attaching metal electrodes to crystals \cite{Bochmann2013}, or by using an interdigital transducer (IDT) to convert an applied electrical voltage into a surface acoustic wave \cite{Balram2016}.
Recently, percent-level transduction efficiencies ($\eta\sim 10^{-2}\textrm{--}10^{-3}$) have been demonstrated in several studies \cite{Jiang2023,Weaver2024,vanThiel2025,Zhao2025}.
Notably, of all the transduction methods, the first realization of the quantum transduction from superconducting qubit to optical photon was done with a GHz piezo-optomechanical crystal in 2020 \cite{Mirhosseini2020}.
The transduction efficiency including a superconducting qubit in the GHz piezo-optomechanical system has been improved from $8.8\times 10^{-6}$ \cite{Mirhosseini2020} to $3.3\times 10^{-3}$ \cite{vanThiel2025}.

\begin{figure*}[!t]
\centering
\includegraphics[width=2\columnwidth]{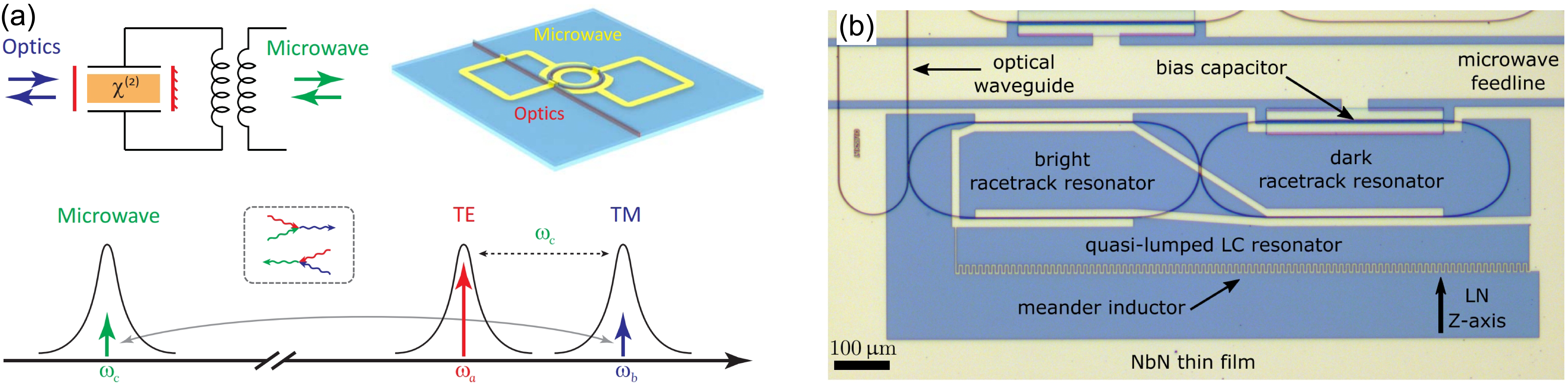}
\caption{{\bf Electro-optic transducers.}
(a) Schematic illustrations of the experimental setup of an electro-optic transducer system.
Adapted from Ref.~\onlinecite{Fan2018}.
Copyright 2018 The Authors, licensed under a Creative Commons Attribution (CC BY) license.
(b) Optical micrograph of an electro-optic transducer device.
Adapted with permission from Ref.~\onlinecite{Holzgrafe2020}.
Copyright 2020 Optical Society of America.
}\label{Fig-Electro-Optic}
\end{figure*}
As we have seen in Eqs.~(\ref{Transduction-bandwidth-one-stage}) and (\ref{Transduction-bandwidth-zero-stage}), the order of the bandwidth in optomechanical systems is determined by the largest among the mechanical, electromechanical, and optomechanical loss rates.
In other words, when the cooperativities $C_{\mathrm{em}}$ and $C_{\mathrm{om}}$ are much smaller (larger) than unity, the mechanical loss rate is dominant (the electromechanical and optomechanical loss rates are dominant).
Owing to the high mechanical frequencies in the GHz regime, the mechanical loss rates are typically of $\mathcal{O}(1\, \mathrm{MHz})$ which is much larger than those of the MHz membranes which are typically of $\mathcal{O}(1\, \mathrm{Hz})$.
Therefore, a high transduction bandwidth of $\mathcal{O}(1\, \mathrm{MHz})$ is expected in the GHz optomechanical systems and indeed has been observed in experiments.

Recently, the use of a nanobeam structure with GHz phonons has been extended to non-piezoelectric systems such as the silicon-on-insulator platform, in which the coupling between phonons and microwave photons is realized via a capacitor with mechanically moving electrodes connected to an external dc voltage source \cite{Zhao2023,Bozkurt2023,Zhao2025}.
This coupling mechanism is similar to that of the MHz electro-optomechanical system with membrane structures in the sense that the phonon mode originates from the quantized mechanical displacement operator $\hat{x}$.
Note, however, that the coupling is resonant, not parametric with microwave pumping (i.e., not enhanced by the factor $\sqrt{n_{\mathrm{e}}}$).
In Ref.~\onlinecite{Zhao2025}, a high transduction efficiency of $\eta=2.2\times 10^{-2}$ as well as a low added noise of $0.94$ has been demonstrated.

Several geometrical structures other than the nanobeam structure have been implemented in the GHz optomechanical quantum transducers \cite{Han2020,Yoon2023,Blesin2024,Chen2023}, which are summarized in Table~\label{Table-MHz-Optomechanical}I.
The thickness mode of a microdisk fabricated from AlN simultaneously supports a high frequency of $\sim 10\, \mathrm{GHz}$ mechanical mode and an optical whispering gallery mode \cite{Han2015,Han2020}.
A superconducting microwave cavity is used as well as an optical cavity in this microdisk system, giving rise to an cavity-enhanced large electromechanical cooperativity $C_{\mathrm{em}}\approx 7.4$ and a transduction efficiency of $\eta=7.3\times 10^{-4}$ \cite{Han2020}.
A microdisk has also been fabricated from diamond \cite{Shandilya2021}.
The bulk acoustic resonators can also host GHz frequency mechanical modes (standing-wave acoustic modes) with a low loss \cite{Galliou2013} and strong optomechanical coupling can be realized via the Brillouin scattering \cite{Renninger2018,Kharel2019,Yoon2023}.
Figure~\ref{Fig-GHz-Optomechanical}(b) shows a schematic illustration of a transducer device with a GHz bulk acoustic resonator \cite{Yoon2023}.
A recent study with Si$_3$N$_4$ bulk acoustic resonator has shown the microwave-to-optical quantum transduction with a transduction efficiency of $\eta=1.6\times 10^{-5}$ and a bandwidth of $25\, \mathrm{MHz}$
 \cite{Blesin2024}.
Another study has demonstrated the microwave-to-optical quantum transduction with an optomechanical ring resonator \cite{Chen2023}.

\begingroup
\renewcommand{\arraystretch}{1.3}
\begin{table*}[!t]
\caption{{\bf Schematic comparison of the microwave-to-optical quantum transduction via the electro-optic effect.}
NR, not reported.
--, not applicable.
$^*$ indicates the utilization of bulk LiNbO$_3$ with a whispering gallery mode resonator.
$^\dag$ indicates the utilization of thin-film LiNbO$_3$ integrated with a superconducting microwave resonator.
$^\clubsuit$ indicates a transduction from superconducting qubit to optical photon.
The efficiency $\eta$ denotes the total transduction efficiency [such as the one given by Eq.~(\ref{Transduction-efficiency-zero-stage-transduction})] except for the transduction from superconducting qubit to optical photon.}
\begin{ruledtabular}
\begin{tabular}{ccccccccc}
Reference &  Ref.~\onlinecite{Rueda2016} & Ref.~\onlinecite{Fan2018} & Ref.~\onlinecite{Holzgrafe2020} & Ref.~\onlinecite{McKenna2020} & Ref.~\onlinecite{Hease2020} & Ref.~\onlinecite{Sahu2022} & Ref.~\onlinecite{Arnold2025}$^\clubsuit$ & Ref.~\onlinecite{Warner2025}$^\clubsuit$\\
(Year) & (2016) & (2018) & (2020) & (2020) & (2020) & (2022) & (2025) & (2025)\\
\hline
System & LiNbO$_3$$^*$ & AlN & LiNbO$_3$$^\dag$ & LiNbO$_3$$^\dag$ & LiNbO$_3$$^*$ & LiNbO$_3$$^*$ & LiNbO$_3$$^*$ & LiNbO$_3$$^\dag$\\
Efficiency $\eta$ & $1.1\times 10^{-3}$ & $2\times 10^{-2}$ & $2.7\times 10^{-5}$ & $6.6\times 10^{-6}$ & $3\times 10^{-4}$ & $0.15$ & $3\times 10^{-3}$ & $1.18\times 10^{-2}$\\
Cooperativity $C_{\mathrm{eo}}$ & $4\times 10^{-3}$ & 0.073 & NR & NR & $1.67\times 10^{-3}$ & 0.38 & NR & $1.16\times 10^{-2}$ \\
Added noise & NR & NR & NR & NR & 21.3 & 0.41 & NR & $< 0.12$ \\
Bandwidth & $\sim 1\, \mathrm{MHz}$ & $0.59\, \mathrm{MHz}$ & $13\, \mathrm{MHz}$ & $20\, \mathrm{MHz}$ & $10.7\, \mathrm{MHz}$ & $24\, \mathrm{MHz}$ & $\sim 10\, \mathrm{MHz}$ & $30\, \mathrm{MHz}$\\
Temperature & $300\, \mathrm{K}$ & $2\, \mathrm{K}$ & $1\, \mathrm{K}$ & $1\, \mathrm{K}$ & $7\, \mathrm{mK}$ & $10\, \mathrm{mK}$ & $10\, \mathrm{mK}$ &  $14\, \mathrm{mK}$\\
\end{tabular}
\end{ruledtabular}
\label{Table-Electro-optic}
\end{table*}
\endgroup
%
\subsection*{Transduction via the electro-optic effect}
The quantum transduction utilizing a linear electro-optic effect, the Pockels effect, has also attracted much attention as a promising approach toward high-efficiency and low-noise transduction.
Since the Pockels effect directly mediates the interaction between microwave and optical photons, the use of the intermediate bosonic mode is not required  \cite{Tsang2010,Tsang2011}.
This means that the generation of the thermal noise and the limitation of the transduction bandwidth due to the intermediate bosonic modes can be avoided.
The low fabrication complexity with scalability is another attractive feature of the electro-optic transducers, since they do not rely on free-standing structures such as membranes and nanobeams.
On the other hand, the electro-optic transducers generically require strong optical pump powers due to the small single-photon coupling strength, which can cause increased noise and heating.

In order to derive the coupling between microwave and optical photons, we start with the fact that the electric polarization can in general be expanded in powers of the applied electric field $\bm{E}$ as
\begin{align}
P_i=\varepsilon_0\left(\sum_j \chi_{ij}^{(1)}E_j+\sum_{j,k}\chi_{ijk}^{(2)}E_jE_k+\sum_{j,k,l}\chi_{ijkl}^{(3)}E_jE_kE_l+\cdots\right),
\end{align}
where $\varepsilon_0$ is the vacuum permittivity, $\chi^{(n)}$ is the $n$-th order susceptibility tensor of rank $(n+1)$, and the subscripts indicate spatial directions.
The Pockels effect is characterized by the second-order susceptibility $\chi^{(2)}_{ijk}\propto r_{hk}$ and occurs in materials without inversion symmetry such as LiNbO$_3$ \cite{Thomaschewski2022}.
Here, $r_{hk}$ is the electro-optic (Pockels) tensor in the contracted Voigt notation.
Now we consider the optically modulated electric polarization $\bm{P}^{(2)}$ due to the Pockels effect, which is quadratic in the optical photon operator.
The interaction Hamiltonian between optical and microwave photons is then given by $H_{\mathrm{EO}}=\int dV\, \bm{P}^{(2)}\cdot\bm{E}$, where $\bm{E}$ is the applied electric field that is responsible for microwave photons.

For concreteness, here we consider the triply resonant interaction between microwave, optical pump, and optical signal photons in their whispering gallery modes \cite{Rueda2016,Javerzac-Galy2016,Hease2020,Sahu2022,Arnold2025}.
In this case, the interaction Hamiltonian reads 
\begin{align}
H_{\mathrm{EO}}=\hbar g_{\mathrm{EO}}\left(\hat{a}_{\mathrm{e}}\hat{a}_{\mathrm{p}}\hat{a}_{\mathrm{o}}^\dag+\hat{a}_{\mathrm{e}}^\dag\hat{a}_{\mathrm{p}}^\dag\hat{a}_{\mathrm{o}}\right),
\label{Hamiltonian-triply-resonant}
\end{align}
where $\hat{a}_{\mathrm{e}}$, $\hat{a}_{\mathrm{p}}$, and $\hat{a}_{\mathrm{o}}$ are the annihilation operators for the microwave, optical pump, and optical signal mode, respectively.
Strongly driving the optical pump mode to the coherent mode allows us to have $\hat{a}_{\mathrm{p}}\to \sqrt{n_{\mathrm{p}}}$ with $n_{\mathrm{p}}$ the pump photon number.
Then, the interaction Hamiltonian reduces to a simpler form, 
\begin{align}
H_{\mathrm{EO}}=\hbar G_{\mathrm{EO}}\left(\hat{a}_{\mathrm{e}}\hat{a}_{\mathrm{o}}^\dag+\hat{a}_{\mathrm{e}}^\dag\hat{a}_{\mathrm{o}}\right),
\label{Hamiltonian-Cavity-enhanced-EO}
\end{align}
with the cavity-enhanced effective coupling strength $G_{\mathrm{EO}}=g_{\mathrm{EO}}\sqrt{n_{\mathrm{p}}}$.
Here, the single-photon coupling strength $g_{\mathrm{EO}}$ is given by $g_{\mathrm{EO}}=r\sqrt{\frac{\varepsilon_{\mathrm{p}}\varepsilon_{\mathrm{o}}}{\varepsilon_{\mathrm{e}}}}\sqrt{ \frac{\hbar\omega_{\mathrm{e}}\omega_{\mathrm{p}}\omega_{\mathrm{o}}}{8\varepsilon_0V_{\mathrm{e}}V_{\mathrm{p}}V_{\mathrm{o}}} }\int d^3x\, \psi_{\mathrm{e}}\psi_{\mathrm{p}}\psi_{\mathrm{o}}^*$ \cite{Rueda2016,Hease2020,Strekalov2016}.
Here, $r$ is a material-dependent $\chi^{(2)}$ coefficient, $\varepsilon_\mu$ is the relative permittivity, $V_\mu$ is the effective mode volume, $\omega_\mu$ is the mode frequency, and $\psi_\mu$ is the spatial distribution function of the photon mode $\mu=\mathrm{e,p,o}$.

The triply resonant systems described by the Hamiltonian of the form of Eq.~(\ref{Hamiltonian-triply-resonant}) can also be realized in various ways.
As shown in Fig.~\ref{Fig-Electro-Optic}(a), we can utilize the transverse-electric (TE) and transverse-magnetic (TM) optical modes as pump and signal modes, respectively, in AlN integrated with a superconducting microwave resonator \cite{Fan2018}.
In this approach, the device size and mode volumes can be reduced since there is no limitation imposed by the free spectral range unlike the case with a whispering gallery mode resonator.
We can also utilize the TE polarized modes exhibiting a large $\chi^{(2)}$ coefficient $r_{33}$ of $30\, \mathrm{pm/V}$ in thin-film LiNbO$_3$ integrated with a superconducting microwave resonator \cite{Soltani2017,Holzgrafe2020,McKenna2020,Xu2021,Warner2025} (see, for example,  Ref.~\onlinecite{McKenna2020} for a detailed theoretical description).

Table~\ref{Table-Electro-optic} summarizes the experimental progress of the microwave-to-optical quantum transduction utilizing the electro-optic effect \cite{Rueda2016,Fan2018,Holzgrafe2020,McKenna2020,Hease2020,Sahu2022,Arnold2025,Warner2025}.
Because of the use of both microwave and optical cavities, the transduction efficiency is given by Eq.~(\ref{Transduction-efficiency-zero-stage-transduction}).
Figure~\ref{Fig-Electro-Optic}(b) shows an optical micrograph of an electro-optic transducer device \cite{Holzgrafe2020}.
As expected from the absence of the intermediate bosonic mode, the transduction bandwidths of the electro-optic transducers are of $\mathcal{O}(10\, \mathrm{MHz})$, which is about an order of magnitude higher than those of the GHz optomechanical transducers.
Here, recall that the transduction bandwidth is given by $\Delta\omega=\kappa_{\mathrm{e}}(1+C_{\mathrm{eo}})$ in the (typical) case of $\kappa_{\mathrm{e}}<\kappa_{\mathrm{o}}$ [see Eq.~(\ref{Transduction-bandwidth-zero-stage})].
On the other hand, the single-photon coupling rate $g_{\mathrm{EO}}/2\pi$ is typically of $\mathcal{O}(1\, \mathrm{kHz})$ \cite{Han2021}, which is rather small compared to the single-photon optomechanical coupling rate of $\mathcal{O}(0.1\textrm{--}1\, \mathrm{MHz})$ in the GHz optomechanical systems.
This implies that in general the electro-optic transducers require high input optical pump powers in order to obtain a high efficiency, which can lead to increased noise and heating.
A generic strategy for improving the single-photon coupling rate $g_{\mathrm{EO}}/2\pi$ is to choose materials with large $\chi^{(2)}$ coefficient and to reduce the effective mode volumes [see Eq.~(\ref{Hamiltonian-Cavity-enhanced-EO})].
Notably, recent studies have demonstrated the quantum transduction from superconducting qubit to optical photon with percent-level transduction efficiencies ($\sim 10^{-2}\textrm{--}10^{-3}$) \cite{Arnold2025,Warner2025}.
In particular, Ref.~\onlinecite{Warner2025} has demonstrated the transduction with a low added noise of $<0.12$ and a high bandwidth of $30\, \mathrm{MHz}$ at an input optical pump power of $44\, \mathrm{\mu W}$ even with the inclusion of a superconducting qubit.

\begin{figure*}[!t]
\centering
\includegraphics[width=1.9\columnwidth]{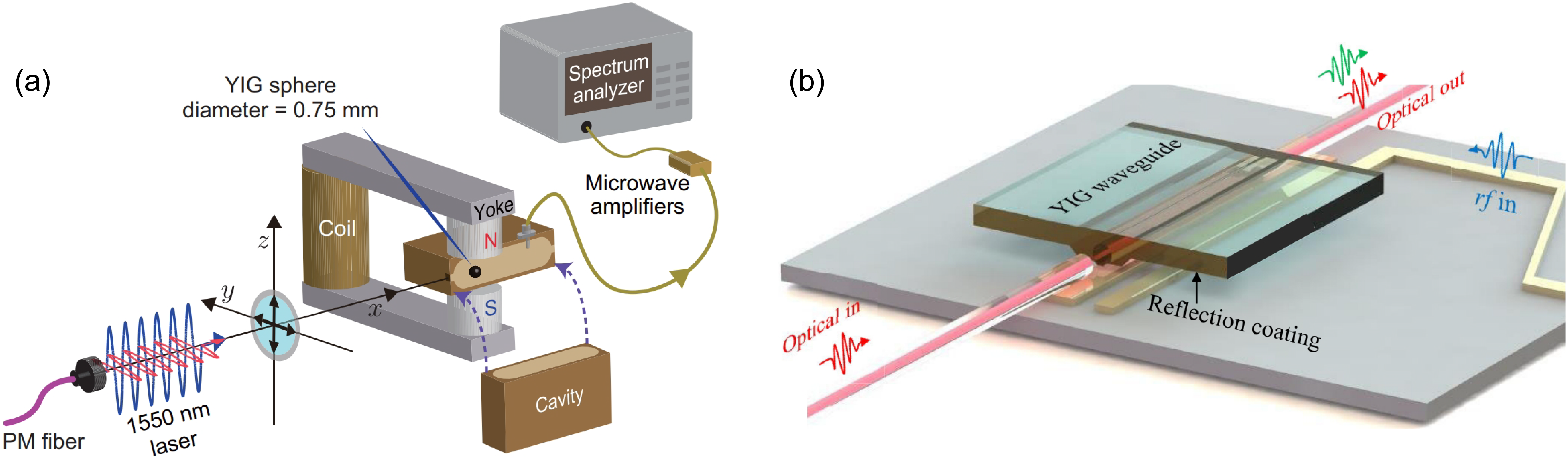}
\caption{{\bf Magneto-optic transducers.}
(a) Schematic illustration of the experimental setup for a quantum transduction via the magneto-optic Faraday effect.
Adapted with permission from Ref.~\onlinecite{Hisatomi2016}.
Copyright 2016 American Physical Society.
(b) Schematic illustration of an integrated optomagnonic waveguide device.
Adapted with permission from Ref.~\onlinecite{Zhu2020}.
Copyright 2020 Optical Society of America.
}\label{Fig-Magneto-optic}
\end{figure*}
%
\subsection*{Transduction via the magneto-optic effect}
The magnons can also serve as the intermediate bosonic mode in a transduction process.
One of the characteristics of magnons is the tunability of the magnon frequency by an external magnetic field, which is in contrast to the optomechanical crystals where the mechanical resonance frequency is basically determined by the design of the structure geometry such as the the size of the holes.
This tunability can be a merit of the magnon-mediated transduction, allowing flexible design of a superconducting qubit.
On the other hand, the magneto-optic transducers currently suffer from the low transduction efficiency due to the weak magneto-optic coupling (i.e., weak light-magnon interaction).

The coupling between magnons and microwave photons comes from the Zeeman interaction between spins $\bm{S}_i$ and the ac magnetic field $\bm{B}$ of the form $H_{\mathrm{Zeeman}}=|\gamma|\sum_i\bm{B}\cdot\bm{S}_i$, where $\gamma$ is the gyromagnetic ratio.
In the presence of a microwave cavity, the interaction Hamiltonian between magnons and microwave cavity photons reads \cite{Tabuchi2014,Zhang2014}
\begin{align}
H_{\mathrm{MO}}=\hbar g_{\mathrm{em}}\left(\hat{a}_{\mathrm{e}}^\dag\hat{a}_{\mathrm{m}}+\hat{a}_{\mathrm{e}}\hat{a}_{\mathrm{m}}^\dag\right),
\end{align}
where $\hat{a}_{\mathrm{e}}$ and $\hat{a}_{\mathrm{m}}$ is the annihilation operators for the microwave cavity mode and the magnon mode, respectively.
The coupling strength $g_{\mathrm{em}}$ is given by $g_{\mathrm{em}}=g_0\sqrt{N_{\mathrm{s}}}$, where $g_0=(|\gamma|/2)\sqrt{\hbar \omega_{\mathrm{e}}\mu_0/V_{\mathrm{c}}}$ 
 is the single-spin interaction strength and $N_{\mathrm{s}}$ is the total spin number in the ferromagnet.
Here, $\omega_{\mathrm{e}}$ is the microwave cavity resonance frequency, $\mu_0$ is the vacuum permeability, and $V_{\mathrm{c}}$ is the cavity mode volume.

\begingroup
\renewcommand{\arraystretch}{1.3}
\begin{table}[!t]
\caption{{\bf Schematic comparison of the microwave-to-optical quantum transduction via the magneto-optic effect.}
NR, not reported.
--, not applicable.
The efficiency $\eta$ denotes the total transduction efficiency [such as the one given by Eq.~(\ref{Transduction-efficiency-one-stage-transduction}) or Eq.~(\ref{Transduction-efficiency-zero-stage-transduction-2})].}
\begin{ruledtabular}
\begin{tabular}{ccccc}
Reference &  Ref.~\onlinecite{Hisatomi2016} & Ref.~\onlinecite{Osada2016} & Ref.~\onlinecite{Zhang2016} & Ref.~\onlinecite{Zhu2020}\\
(Year) & (2016) & (2016) & (2016) & (2020)\\
\hline
System & YIG & YIG & YIG & YIG\\
Efficiency $\eta$ & $\sim 10^{-10}$ & $7\times10^{-14}$ & $1.7\times 10^{-15}$ & $1.1\times 10^{-8}$\\
Cooperativity $C_{\mathrm{em}}$ & 510 & -- & -- & 0.87\\
Cooperativity $C_{\mathrm{om}}$ & -- & NR & $5.4\times 10^{-7}$ & $4.1\times 10^{-7}$\\
Bandwidth & NR & NR & NR & $16.1\, \mathrm{MHz}$ \\
Temperature & $\sim 300\, \mathrm{K}$ & $\sim 300\, \mathrm{K}$ & $\sim 300\, \mathrm{K}$ & $\sim 300\, \mathrm{K}$\\
\end{tabular}
\end{ruledtabular}
\label{Table-Magneto-optic}
\end{table}
\endgroup
The coupling between magnons and optical photons comes from a linear magneto-optic effect known as the Faraday effect, which is described by the Hamiltonian \cite{Landau-book} $H_{\mathrm{MO}}=-i(\varepsilon_0 f/4)\int d^3r\, \bm{M}(\bm{r})\cdot[\bm{E}^*(\bm{r})\times\bm{E}(\bm{r})]$, where $\bm{M}(\bm{r})$ is the magnetization and $f\, (\propto \theta_{\mathrm{F}})$ is a material-dependent constant that is related to the Faraday rotation angle per unit length $\theta_{\mathrm{F}}$.
Introducing quantized electric fields, the interaction Hamiltonian is generically written as
\begin{align}
H_{\mathrm{MO}}=\sum_{\alpha,\beta}\hbar g_{\mathrm{MO},\alpha\beta}\left(\hat{a}_{\mathrm{m}}+\hat{a}_{\mathrm{m}}^\dag\right)\hat{a}_{\alpha}^\dag\hat{a}_{\beta},
\end{align}
where $\hat{a}_{\alpha(\beta)}$ is the annihilation operator for the optical mode $\alpha\, (\beta)$.
In the presence of an optical cavity (such as the whispering gallery mode resonator) with the steady-state pump photon number $\bar{n}_{\mathrm{cav}}=|\langle \hat{a}_{\mathrm{p}}\rangle|^2$, the interaction Hamiltonian between optical signal photons and magnons reads \cite{Osada2016,Zhang2016,ViolaKusminskiy2016}
\begin{align}
H_{\mathrm{MO}}=\hbar G_{\mathrm{MO}}\left(\hat{a}_{\mathrm{o}}^\dag\hat{a}_{\mathrm{m}}+\hat{a}_{\mathrm{o}}\hat{a}_{\mathrm{m}}^\dag\right),
\label{Hamiltonian-Faraday-effect}
\end{align}
where $\hat{a}_{\mathrm{o}}$ is the annihilation operator for the optical signal mode.
Here, the coupling strength $G_{\mathrm{MO}}$ is given by $G_{\mathrm{MO}}=g_{\mathrm{MO},0}\sqrt{\bar{n}_{\mathrm{cav}}}$, where  $g_{\mathrm{MO},0}=c\theta_{\mathrm{F}}/(4\sqrt{2\varepsilon_{\mathrm{r}} N_{\mathrm{s}}})$ with $c$ the speed of light, $\theta_{\mathrm{F}}$ the Faraday rotation angle per unit length, and $\varepsilon_{\mathrm{r}}$ the relative permittivity.

Table~\ref{Table-Magneto-optic} summarizes the experimental progress of the microwave-to-optical quantum transduction via the magneto-optic effect \cite{Hisatomi2016,Osada2016,Zhang2016,Zhu2020}.
So far, the quantum transduction has been demonstrated using the ferrimagnetic insulator YIG (yttrium iron garnet) at room temperature.
The YIG is known for a low magnon decay rate of $\kappa_{\mathrm{m}}\approx 1\, \mathrm{MHz}$ \cite{Tabuchi2014,Zhang2014,Rameshti2022} with the ferromagnetic resonance frequency $\omega_{\mathrm{m}}\sim 10\, \mathrm{GHz}$ which is tunable by an external static magnetic field.
The quantum transduction with YIG has so far been demonstrated in the case with a microwave cavity but without an optical cavity \cite{Hisatomi2016}, in the case without a microwave cavity but with an optical cavity \cite{Osada2016,Zhang2016}, and in the case with microwave and optical cavities \cite{Zhu2020}.
In the former two cases (the last case), the transduction efficiency $\eta$ is given by Eq.~(\ref{Transduction-efficiency-zero-stage-transduction-2}) [Eq.~(\ref{Transduction-efficiency-one-stage-transduction})].
Figure~\ref{Fig-Magneto-optic} shows schematic illustrations of the experimental setup of a microwave-to-optical transduction without an optical cavity \cite{Hisatomi2016} and an integrated optomagnonic waveguide device \cite{Zhu2020}.
Note that the order of magnitude of a high bandwidth $16.1\, \mathrm{MHz}$ obtained in Ref.~\onlinecite{Zhu2020} would be understood from a generic expression that is applicable to the one-stage transduction, $\Delta\omega=\kappa_{\mathrm{m}}(1+C_{\mathrm{em}}+C_{\mathrm{om}})$ [Eq.~(\ref{Transduction-bandwidth-one-stage})], with $\kappa_{\mathrm{m}}/2\pi=3.25\, \mathrm{MHz}$, $C_{\mathrm{em}}=0.87$, and  $C_{\mathrm{om}}=4.1\times 10^{-7}$.

\begin{figure*}[!t]
\centering
\includegraphics[width=2\columnwidth]{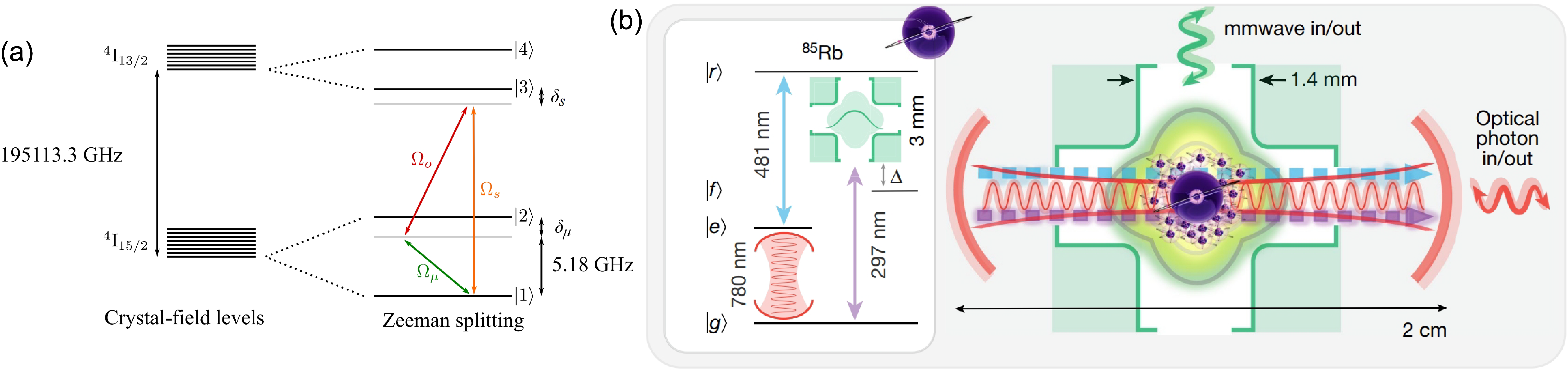}
\caption{{\bf Transducers with atomic ensembles.}
(a) Energy-level diagram of a three-level system, erbium-doped Y$_2$SiO$_5$.
Adapted with permission from Ref.~\onlinecite{Fernandez-Gonzalvo2019}.
Copyright 2019 American Physical Society.
(b) Schematic illustration of a transduction system with Rydberg atoms.
Adapted with permission from Ref.~\onlinecite{Kumar2023}.
Copyright 2023 Springer Nature.
}\label{Fig-Atomic-ensembles}
\end{figure*}
The low transduction efficiency in YIG is due to the weakness of the coupling between magnons and optical photons.
As can be seen from Eq.~(\ref{Hamiltonian-Faraday-effect}), the coupling strength between magnons and optical photons is proportional to $\theta_{\mathrm{F}}/\sqrt{N_{\mathrm{s}}}$, which means that the coupling can be enhanced by reducing the volume of the ferromagnet and by choosing a ferromagnet with a large Faraday rotation angle.
Recently, it has been proposed that the transduction efficiency between microwave and terahertz photons can be improved to $\eta\sim 10^{-3}\textrm{--}10^{-4}$ (even without an optical cavity) by utilizing the heterostructures consisting of topological insulator thin films such as Bi$_2$Se$_3$ and ferromagnetic insulator thin films such as YIG \cite{Sekine2024}, while that of YIG alone is $\eta\sim 10^{-8}$ at the same input optical power in the terahertz regime.
The mechanism for this improvement is the topological Faraday effect of topological insulators that is independent
of the sample thickness in the terahertz regime, leading to a large Faraday rotation angle and thus enhanced
light-magnon interaction in the thin-film limit.

Recently, it has also been proposed theoretically that antiferromagnets can also be utilized for the microwave-to-optical quantum transduction \cite{Sekine2025}.
In sharp contrast to the case of ferromagnets, the transduction can occur even at zero applied magnetic field.
Owing to the wide tunability of the antiferromagnetic resonance frequency from $\mathcal{O}(1\, \mathrm{GHz})$ to $\mathcal{O}(1\, \mathrm{THz})$, it is expected that a variety of quantum devices that operate at this frequency range can be interconnected via the quantum transducer utilizing antiferromagnets.

Finally, we note that the coherent coupling between a superconducting qubit and a magnon, which is an essential step toward the realization of the quantum state transfer between superconducting qubits via optical photon, has been experimentally observed \cite{Tabuchi2015,Viennot2015,Lachance-Quirion2020}, although the quantum transduction from a superconducting qubit to optical photons is yet to be realized.

\begingroup
\renewcommand{\arraystretch}{1.3}
\begin{table}[!t]
\caption{{\bf Schematic comparison of the microwave-to-optical quantum transduction via rare-earth ions.}
NR, not reported.
The efficiency $\eta$ denotes the total transduction efficiency [such as the one given by Eq.~(\ref{Transduction-efficiency-zero-stage-transduction})].}
\begin{ruledtabular}
\begin{tabular}{ccccc}
Reference &  Ref.~\onlinecite{Fernandez-Gonzalvo2019} & Ref.~\onlinecite{Bartholomew2020} & Ref.~\onlinecite{Rochman2023} & Ref.~\onlinecite{Xie2025} \\
(Year) & (2019) & (2020) & (2023) & (2025)\\
\hline
System & Er$^{3+}$:Y$_2$SiO$_5$ & Yb$^{3+}$:YVO$_4$ & Er$^{3+}$:YVO$_4$  & Yb$^{3+}$:YVO$_4$\\
Efficiency $\eta$ & $1.26\times 10^{-5}$ & $1.2\times 10^{-13}$ & $1\times 10^{-7}$ & $7.6\times 10^{-3}$ \\
Added noise & NR & NR & NR & 1.24 \\
Bandwidth & NR & $0.1\, \mathrm{MHz}$ & NR & $0.5\, \mathrm{MHz}$ \\
Temperature & $4.6\, \mathrm{K}$ & $\sim 40\, \mathrm{mK}$ & $\sim 100\, \mathrm{mK}$ & $< 1\, \mathrm{K}$\\
\end{tabular}
\end{ruledtabular}
\label{Table-Rare-earth}
\end{table}
\endgroup
%
\subsection*{Transduction with atomic ensembles}
Utilizing atomic ensembles for the quantum transduction is based on the idea such that atomic ensembles naturally have energy-level transitions which can be manipulated by microwave and optical fields.
Therefore, a variety of proposals have been made so far, including ensembles of rare-earth ion dopants, ensembles of  neutral atoms, and color centers in diamond.
In particular, one of the advantages of utilizing the three-level systems such as those of rare-earth dopants is that the intermediate bosonic mode is not required for the microwave-to-optical quantum transduction, as in the case of the transduction via the electro-optic effect.
This implies that the thermal noise associated with the intermediate bosonic mode is absent, and the transduction bandwidth is not limited by the loss rate of the intermediate bosonic mode (which is usually the smallest among the loss rates).
On the other hand, the device size of the ensemble-based transducers tends to be large due to e.g., magnetic field shielding,  posing a challenge for scalable on-chip integration.

\begin{figure*}[!t]
\centering
\includegraphics[width=1.8\columnwidth]{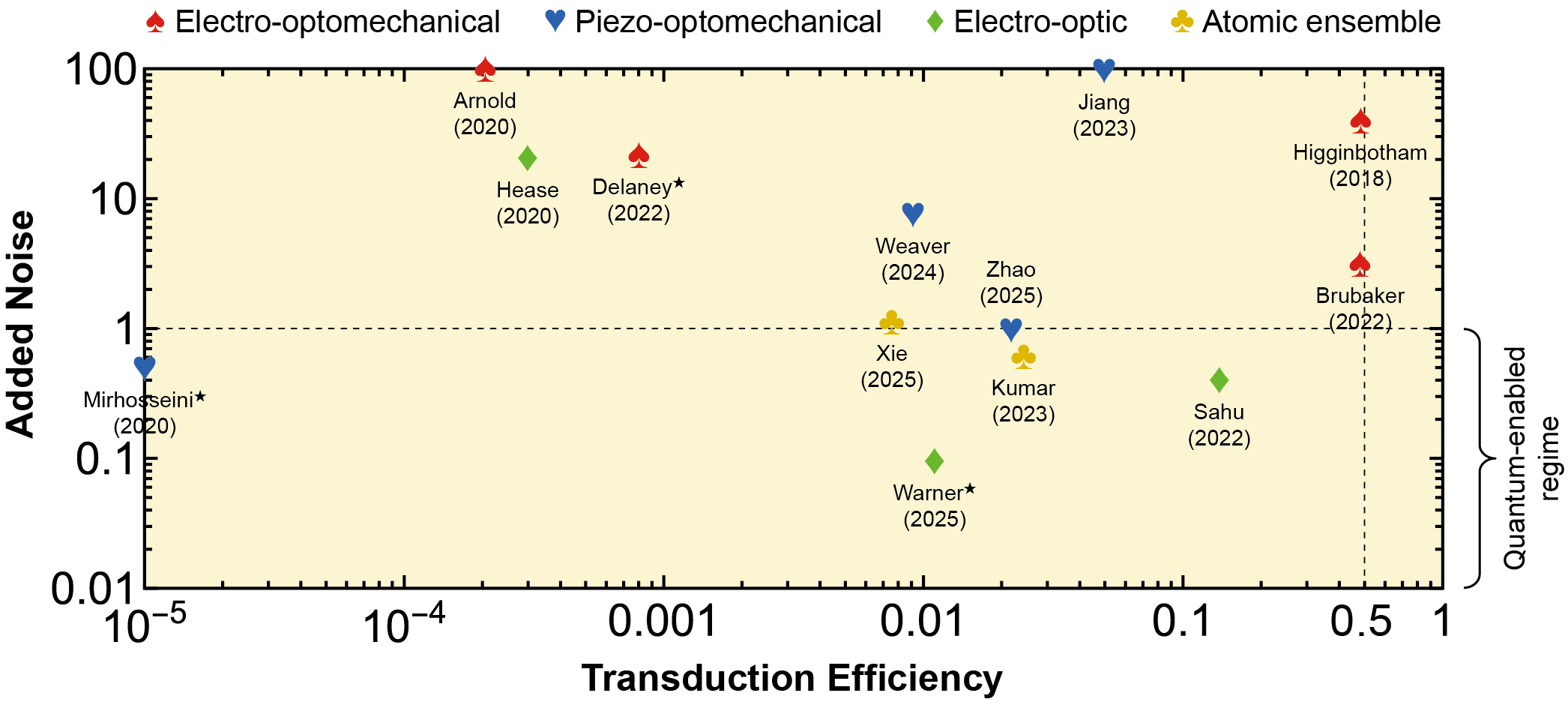}
\caption{{\bf Brief summary of various microwave-to-optical quantum transduction methods in terms of the transduction efficiency and the added noise.}
$^\star$ indicates a transduction from superconducting qubit to optical photon.
The region with $N_{\mathrm{add}}<1$ is usually called the quantum-enabled regime.
The transduction bandwidths are $12\, \mathrm{kHz}$ in Higginbotham (2018) \cite{Higginbotham2018}, $0.37\, \mathrm{kHz}$ in Arnold (2020) \cite{Arnold2020}, $2\, \mathrm{kHz}$ in Brubaker (2022) \cite{Brubaker2022}, $6.1\, \mathrm{kHz}$ in Delaney (2022) \cite{Delaney2022}, $1\, \mathrm{MHz}$ in Mirhosseini (2020) \cite{Mirhosseini2020}, $1.5\, \mathrm{MHz}$ in Jiang (2023) \cite{Jiang2023}, $14.8\, \mathrm{MHz}$ in Weaver (2024) \cite{Weaver2024}, $88.9\, \mathrm{kHz}$ in Zhao (2025) \cite{Zhao2025}, $10.7\, \mathrm{MHz}$ in Hease (2020) \cite{Hease2020}, $24\, \mathrm{MHz}$ in Sahu (2022) \cite{Sahu2022}, $30\, \mathrm{MHz}$ in Warner (2025) \cite{Warner2025}, $0.5\, \mathrm{MHz}$ in Xie (2025) \cite{Xie2025}, and $0.36\, \mathrm{MHz}$ in Kumar (2023) \cite{Kumar2023}.
}\label{Fig-Efficiency-vs-Noise}
\end{figure*}
For concreteness, here we consider an ensemble of three-level atoms such as the $\Lambda$ system and the V system (with the states $|1\rangle$, $|2\rangle$, and $|3\rangle$) interacting with microwave cavity photons and optical cavity photons, as shown in Fig.~\ref{Fig-Atomic-ensembles}(a).
We consider the following Hamiltonian of the system \cite{Williamson2014,Khalifa2025}
\begin{align}
H_{\mathrm{REI}}/\hbar=&\, \sum_i\left[\left(\delta_{3,i}|3\rangle_i\langle 3|_i+\delta_{2,i}|2\rangle_i\langle 2|_i\right)\right.\nonumber\\
&+\left.\left(\Omega_{i}|3\rangle_i\langle 2|_i+g_{\mathrm{e},i}|2\rangle_i\langle 1|_i\hat{a}_{\mathrm{e}}+g_{\mathrm{o},i}|3\rangle_i\langle 1|_i\hat{a}_{\mathrm{o}}+\mathrm{H.c.}\right)\right],
\end{align}
where the summation runs over the active rare-earth ion dopants, $\delta_{3,i}$ ($\delta_{2,i}$) is the detuning from the optical (microwave) cavity frequency, $\Omega_i$ is the Rabi frequency, and $g_{\mathrm{e},i}$ ($g_{\mathrm{o},i}$) is the coupling strength to the microwave (optical) cavity photons.
In the regime of large detunings such that $|\delta_{3,i}|\gg |g_{\mathrm{o},i}|$, $|\delta_{2,i}|\gg |g_{\mathrm{e},i}|$, and $|\delta_{3,i}\delta_{2,i}|\gg |\Omega_i|^2$, the excited states of the atoms $|3\rangle$ and $|2\rangle$ can be adiabatically eliminated \cite{Brion2007}.
As a result we obtain the Hamiltonian describing a linear coupling between microwave cavity and optical cavity photons \cite{Williamson2014}
\begin{align}
H_{\mathrm{REI}}=\hbar \left(G_{\mathrm{REI}}\hat{a}_{\mathrm{e}}\hat{a}_{\mathrm{o}}^\dag+G_{\mathrm{REI}}^*\hat{a}_{\mathrm{e}}^\dag\hat{a}_{\mathrm{o}}\right),
\end{align}
where the coupling strength $G_{\mathrm{REI}}$ is given by $G_{\mathrm{REI}}=\sum_i\Omega_i g_{\mathrm{e},i}g_{\mathrm{o},i}^* /(\delta_{3,i}\delta_{2,i}-|\Omega_i|^2)\approx \sum_i\Omega_i g_{\mathrm{e},i}g_{\mathrm{o},i}^* /(\delta_{3,i}\delta_{2,i})$.
Introducing the total decay rates $\kappa_{\mathrm{e}}$ and $\kappa_{\mathrm{o}}$ of the microwave and optical cavity, respectively, the transduction efficiency is given by the same form as that of the electro-optic transducers, i.e., by Eq.~(\ref{Transduction-efficiency-zero-stage-transduction}) with the cooperativity $C_{\mathrm{eo}}=4|G_{\mathrm{REI}}|^2/(\kappa_{\mathrm{e}}\kappa_{\mathrm{o}})$.

Table~\ref{Table-Rare-earth} summarizes the experimental progress of the microwave-to-optical quantum transduction with rare-earth ions \cite{Fernandez-Gonzalvo2019,Bartholomew2020,Rochman2023,Xie2025}.
So far, the transduction has been demonstrated with erbium ions doped into Y$_2$SiO$_5$ \cite{Fernandez-Gonzalvo2019} and YVO$_4$ \cite{Rochman2023} crystals and with ytterbium ions doped into YVO$_4$ \cite{Bartholomew2020,Xie2025} crystals.
Erbium is attractive because its $^4I_{15/2}$ -- $^4I_{13/2}$ optical transitions occur at $\approx 195\, \mathrm{THz}$ which is the frequency currently used for telecom optical fibers [see Fig.~\ref{Fig-Atomic-ensembles}(a)], and the transitions have narrow inhomogeneous linewidths.
Ytterbium-171 is appearing because it has gigahertz-frequency hyperfine transitions as well as narrow inhomogeneous linewidths, which means that the transduction can be done at zero or nearly-zero magnetic field.
It is notable that, in the latest experiment utilizing ytterbium-171 ions doped into an YVO$_4$
crystal \cite{Xie2025}, a high transduction efficiency of $\eta=7.6\times 10^{-3}$ with a low added noise of 1.24 photons has been achieved.
Significant efforts are underway to integrate atomic ensembles with nanophotonic platforms.
For example, rare-earth-ion-doped crystals have been successfully coupled to nanophotonic cavities.
These on-chip devices, such as those with a photonic crystal cavity \cite{Bartholomew2020} and with planar photonic and superconducting resonators \cite{Rochman2023,Xie2025}, not only reduce the device footprint but also enhance the coupling strength between the fields and the ensemble through strong field confinement in the small mode volume of the cavities.

\begingroup
\renewcommand{\arraystretch}{1.3}
\begin{table*}[!t]
\caption{{\bf Schematic comparison of the quantum transduction from superconducting qubit to optical photon.}
NR, not reported.
$^\dag$ ($^\star$) indicates the use of a frequency-tunable (fixed-frequency) transmon qubit.
``Pump power'' refers to the optical input pump power.
See the main text for the definitions of ``Efficiency''.
}
\begin{ruledtabular}
\begin{tabular}{ccccccc}
Transduction method & Efficiency & Added noise & Bandwidth & Pump power & Operating temp. & SC qubit type \\
\hline
Electro-optomechanical effect~\cite{Delaney2022} & $8\times 10^{-4}$ & 23 & $6.1\, \mathrm{kHz}$ & NR & 40$\, \mathrm{mK}$ & Transmon $5.63\, \mathrm{GHz}$ \\
Piezo-optomechanical effect~\cite{Mirhosseini2020} & $8.8\times 10^{-6}$ & 0.57 & $1\, \mathrm{MHz}$ & $2\, \mathrm{\mu W}$ & 15$\, \mathrm{mK}$ & Transmon $5.12\, \mathrm{GHz}$$^\dag$ \\
Piezo-optomechanical effect~\cite{vanThiel2025} & $3.3\times 10^{-3}$ & $2\times 10^3$ & $4.7\, \mathrm{MHz}$ & $31\, \mathrm{\mu W}$ & 25$\, \mathrm{mK}$ & Transmon $4.07\, \mathrm{GHz}$$^\star$  \\
Electro-optic effect~\cite{Arnold2025} & $3\times 10^{-3}$ & NR & $\sim 10\, \mathrm{MHz}$ & $\sim 100\, \mathrm{mW}$ & 10$\, \mathrm {mK}$ & Transmon $6.25\, \mathrm{GHz}$$^\star$ \\
Electro-optic effect~\cite{Warner2025} & $1.18\times 10^{-2}$ & $<0.12$ & $30\, \mathrm{MHz}$ & $44\, \mathrm{\mu W}$ & 14$\, \mathrm {mK}$ &  Transmon $3.70\, \mathrm{GHz}$$^\dag$ \\
\end{tabular}
\end{ruledtabular}
\label{Table-Transduction-with-qubit}
\end{table*}
\endgroup
The microwave-to-optical quantum transduction has also been demonstrated using Rydberg atoms.
So far, the transduction has been demonstrated using six-wave mixing in cold $^{87}$Rb atoms \cite{Han2018,Vogt2019}, three-wave mixing in hot $^{85}$Rb atoms at room temperature \cite{Adwaith2019}, six-wave mixing in hot $^{85}$Rb atoms at room temperature \cite{Tu2022}, and four-wave mixing in cold $^{85}$Rb atoms \cite{Kumar2023}.
Figure~\ref{Fig-Atomic-ensembles}(b) shows a schematic illustration of a transduction system \cite{Kumar2023}.
We note that, in the case of the transduction with Rydberg atoms, the microwave frequency can be as large as $\sim 50\textrm{--}100\, \mathrm{GHz}$ \cite{Han2018,Vogt2019,Tu2022,Kumar2023}.
An efficient quantum transduction with the transduction efficiency of $2.5\times 10^{-2}$ (internal efficiency of $58\%$), bandwidth of $0.36\, \mathrm{MHz}$, and added noise of $0.6$ photons has been observed \cite{Kumar2023}.
Also, in Ref.~\onlinecite{Tu2022} a transduction efficiency of $0.82$ with a bandwidth of about $1\, \mathrm{MHz}$ was achieved, which to our knowledge would be the highest value of all transduction methods.
It should be noted that in the quantum transduction with Rydberg atoms the transduction efficiency is defined by \cite{Han2018,Vogt2019,Adwaith2019,Tu2022}
\begin{align}
\eta=\frac{P_L/(\hbar\omega_L)}{I_M S_M/(\hbar\omega_M)},
\end{align}
where $P_{L}$ is the power of the optical field, $I_M$ is the power intensity of the microwave field, $\omega_{L(M)}$ is the frequency of the optical (microwave) field, and $S_M$ is the cross section of the atomic could.
This definition differs from the conventional definition we have seen in Eqs.~(\ref{Transduction-efficiency-one-stage-transduction}) and (\ref{Transduction-efficiency-zero-stage-transduction}).
On the other hand, in Ref.~\onlinecite{Kumar2023} an expression for the transduction efficiency that is written in terms of cooperativities but takes a different form from Eq.~(\ref{Transduction-efficiency-one-stage-transduction}) have been introduced.

Finally, we note that the coherent coupling between a superconducting qubit and an atomic ensemble, which is an essential step toward the realization of the quantum state transfer between superconducting qubits via optical photon, has been experimentally observed in spin ensembles of the nitrogen-vacancy centers in diamond \cite{Zhu2011,Kubo2011,Saito2013}, although the quantum transduction from a superconducting qubit to optical photons is yet to be realized.

\subsection*{Brief summary of various transduction methods}
So far, we have overviewed the experimental progresses of various methods for the microwave-to-optical transduction on a one-by-one basis, showing explicitly the physical origins of the interaction between microwave and optical photons or the interaction between photons and the intermediate bosonic modes.
Figure~\ref{Fig-Efficiency-vs-Noise} shows a brief summary of the transduction methods in terms of the transduction efficiency $\eta$ and the added noise $N_{\mathrm{add}}$.
From the viewpoint of the high transduction efficiency, the best efficiency observed so far is $\eta=0.47$ with the MHz electro-optomechanical effect \cite{Higginbotham2018,Brubaker2022}.
Also, it is notable that a transduction from superconduting qubit to optical photon with a high efficiency of $\eta=0.19$ (and with $N_{\mathrm{add}}=23$) has been demonstrated with the MHz electro-optomechanical effect \cite{Delaney2022}.
On the other hand, from the view point of the low added noise, several studies have reached or gotten close to the so-called quantum-enabled regime of $N_{\mathrm{add}}<1$.
With the GHz piezo-optomechanical effect, low added noise of $N_{\mathrm{add}}=0.57$ \cite{Mirhosseini2020} and $N_{\mathrm{add}}=0.94$ \cite{Zhao2025} have been demonstrated.
With the electro-optic effect, low added noise of $N_{\mathrm{add}}=1.1$ \cite{Hease2020}, $N_{\mathrm{add}}=0.41$ \cite{Sahu2022}, and $N_{\mathrm{add}}<0.12$ \cite{Warner2025} have been demonstrated.
With atomic ensembles, low added noise of $N_{\mathrm{add}}=1.24$ \cite{Xie2025} and $N_{\mathrm{add}}=0.6$ \cite{Kumar2023} have been demonstrated.
In particular, it is attractive that a transduction from superconducting qubit to optical photon with a high efficiency of $\eta=1.18\times 10^{-2}$ and a low added noise of $N_{\mathrm{add}}<0.12$ has been simultaneously demonstrated with the electro-optic effect \cite{Warner2025}.
However, the region with $\eta>1/2$ and $N_{\mathrm{add}}\ll 1$, where the quantum state transfer between distant superconducting qubits over optical fibers is enabled, is yet to be reached.
We also note that the highest transduction efficiency of $0.82$ with a bandwidth of about $1\, \mathrm{MHz}$ has been observed using Rydberg atoms at room temperature \cite{Tu2022}, although this work is not plotted in Fig.~\ref{Fig-Efficiency-vs-Noise}.

\begin{figure*}[!t]
\centering
\includegraphics[width=2\columnwidth]{Fig-Qubit-to-Optical.png}
\caption{{\bf Quantum transduction from superconducting qubit to optical photon.}
(a) Schematic illustration and numerical simulation of the quantum transduction process utilizing the piezo-optomechanical effect.
Adapted with permission from Ref.~\onlinecite{Mirhosseini2020}.
Copyright 2020 Springer Nature.
(b) Optical micrograph of the fabricated device corresponding to (a).
Adapted with permission from Ref.~\onlinecite{Mirhosseini2020}.
Copyright 2020 Springer Nature.
(c) Optically driven Rabi oscillation of a superconducting qubit.
Adapted with permission from Ref.~\onlinecite{Warner2025}.
Copyright 2025 Springer Nature.
(d) Schematic illustration of dilution refrigerators for a typical current superconducting qubit system with microwave control and microwave readout (left), a possible future one with microwave control and optical readout enabled by a quantum transducer (middle), and a possible future one with optical control and optical readout enabled by a quantum transducer (right).
Adapted from Ref.~\onlinecite{Arnold2025}.
Copyright 2025 The Authors, licensed under a Creative Commons Attribution (CC BY) license.
}\label{Fig-Qubit-to-Optical}
\end{figure*}
%
\section*{Transduction from superconducting qubit to optical photon}
In this section, we overview recent experimental progress on the quantum transduction from superconducting qubit to optical photon, which is an essential step toward the realization of quantum state transfer and remote entanglement between distant superconducting qubits over optical fibers.

Table~\ref{Table-Transduction-with-qubit} summarizes the experimental progress of the quantum transduction from superconducting qubit to optical photon.
To our knowledge, the transduction from superconducting qubit to optical photon has been experimentally demonstrated firstly with the GHz piezo-optomechanical effect in 2020 \cite{Mirhosseini2020}, and subsequently with the MHz electro-optomechanical effect in 2022 \cite{Delaney2022}, with the GHz piezo-optomechanical effect in 2025 \cite{vanThiel2025}, and with the electro-optic effect in 2025 \cite{Arnold2025,Warner2025}.
In these five works, the superconducting qubits of transmon type are used and the quantum transducers are placed near the superconducting qubits, i.e., at the millikelvin stages.

Figures~\ref{Fig-Qubit-to-Optical}(a) and \ref{Fig-Qubit-to-Optical}(b) show the experimental setup in Ref.~\onlinecite{Mirhosseini2020}, where the quantum Rabi oscillations of a superconducting qubit were observed with a low added noise via single-photon detection of the converted optical photon over an optical fiber.
The overall efficiency and added noise (referred to the qubit) of the transduction process were found to be $\eta=P_\pi-P_0\approx 8.8\times 10^{-5}$ and $N_{\mathrm{add}}=P_0/\eta\approx 0.57$.
Here, $P_\pi$ ($P_0$) is the single optical photon detection probability with (without) a $\pi$-pulse to the qubit, which is calibrated assuming that the qubit acts as a single microwave photon source.
Thus, $P_\pi$ is understood as the number of the emitted (and converted) photons during the Rabi oscillation of the qubit, while $P_0$ is understood as the number of noise photons.

In Refs.~\onlinecite{Delaney2022,vanThiel2025,Arnold2025}, the quantum efficiency is used to characterize the transduction from superconducting qubit to optical photon.
The quantum efficiency can be understood as a figure of merit describing the total performance of the qubit readout through a transducer device, which is defined by $\eta_q=a^2\sigma^2/2$ \cite{Bultink2018,Delaney2022}.
Here, $a$ is the slope of the signal-to-noise ratio of the qubit readout and $\sigma$ is the width of the Gaussian distribution representing the qubit dephasing given by the off-diagonal elements of the qubit's density matrix.
The observed maximum quantum efficiency was $\eta_q\approx 8\times 10^{-4}$ \cite{Delaney2022}.
The quantum efficiency can decomposed as
$\eta_q=\eta_{\mathrm{bw}}\eta_{\mathrm{t}}\eta_{\mathrm{G}}\eta_{\mathrm{mic}}\eta_{\mathrm{opt}}\eta_{\mathrm{cav}}\eta_{\mathrm{noise}}$ \cite{Delaney2022}.
Among these contributions, the transducer efficiency $\eta_{\mathrm{t}}$, which takes the form of Eq.~(\ref{Transduction-efficiency-one-stage-transduction}), was $\eta_{\mathrm{t}}=0.19$.
This is very high and can be understood as a characteristic of the MHz membrane-based transducers.
On the other hand, the added noise at the input of the transducer was $N_{\mathrm{add}}=23$.
In Ref.~\onlinecite{Delaney2022}, the backaction imparted from the transducer, i.e., the number of excess noise photons in the microwave readout cavity dispersively coupled to the qubit was also characterized.
Notably, a low backaction of $\Delta n\approx 3\times 10^{-3}$ photons from the transducer on the qubit was realized, which is an important step toward remote entanglement of qubits as the backaction on the qubit can limit the fidelity of entanglement.

A coherent optical control of a superconducting qubit was demonstrated via the optical-to-microwave transduction in Ref.~\onlinecite{Warner2025}.
As shown in Fig.~\ref{Fig-Qubit-to-Optical}(c), the Rabi oscillations of the qubit was driven by an optical pulse via an electro-optic transducer.
In this study, the transduction efficiency $\eta$ is characterized by comparing the output signal flux with the input idler flux.
The peak efficiency and added microwave noise were found to be $\eta=1.18\times 10^{-2}$ and $N_{\mathrm{add}}<0.12$, which is impressive since the transduction process even includes a superconducting qubit.
Here, on a related note, we note that an optical control and readout of a superconducting qubit has also been demonstrated using a photonic link that is capable of directly delivering microwave signals at millikelvin temperatures with an optical fiber guiding modulated laser light from room temperature to a cryogenic photodetector \cite{Lecocq2021}.
Combining with the latest demonstrations of the optical readout of a superconducting qubit with a fidelity higher than $80\, \%$ \cite{vanThiel2025,Arnold2025}, all-optical control and readout of superconducting qubits via microwave-to-optical quantum transducers may become possible as illustrated in Fig.~\ref{Fig-Qubit-to-Optical}(d).

\section*{Discussion and outlook}
Generically, the transduction efficiency can be improved by increasing the input pump photon number $\bar{n}_{\mathrm{cav}}$ (i.e., the input power $P$), since the cooperativity is proportional to $\bar{n}_{\mathrm{cav}}$.
Here, the pump photon number is given by $\bar{n}_{\mathrm{cav}}\propto P/(\hbar\omega_{\mathrm{p}})$ with $\omega_{\mathrm{p}}$ the pump frequency \cite{Aspelmeyer2014}.
However, as a trade-off, there will be an increase in the environmental temperature.
This problem is crucial because the transducer system should be placed in the millikelvin stage of a dilution refrigerator in order to keep the thermal noise from the microwave domain sufficiently low.
Thus, it is challenging but necessary to achieve a lower pump power while maintaining a high transduction efficiency.
For this, improving the single-photon coupling rate by the material choice and device design is important.
Also, improving the quality factors of the cavity and the intermediate bosonic mode (i.e., obtaining 
lower loss rates) is important, which, however, comes with a trade-off between the transduction bandwidth.
This is because the transduction bandwidth of the one-stage transduction is given by $\Delta\omega=\kappa_{\mathrm{m}}(1+C_{\mathrm{em}}+C_{\mathrm{om}})$, i.e., is determined by the dynamically broadened linewidth of the intermediate bosonic mode, and the transduction bandwidth of the zero-stage transduction (via the electro-optic effect or atomic ensembles) is given by $\Delta\omega=\kappa_{\mathrm{e}}(1+C_{\mathrm{eo}})$, i.e., is determined by the performance (loss rate) of the cavities, where we have assumed that $\kappa_{\mathrm{e}}< \kappa_{\mathrm{o}}$.

Generating non-classical (quantum) microwave-optical photon entanglement via quantum transducers is another important direction toward the realization of remote entanglement between distant superconducting quantum processors \cite{Rueda2019,Wu2021,Sahu2023,Meesala2024a,Meesala2024b}.
These efforts aim for the quantum state transfer via quantum teleportation after establishing the heralded entanglement generation assisted with two-way classical signaling \cite{Duan2001,Muralidharan2016,Zhong2020,Krastanov2021}.
Using a bulk electro-optic transducer, an entanglement between propagating microwave and optical fields in the continuous variable domain has been generated \cite{Sahu2023}.
Also, in a piezo-optomechanical transducer integrated with superconducting circuits, non-classical microwave-optical photon
pair \cite{Meesala2024a} and Bell states between microwave and optical photons \cite{Meesala2024b} have been generated.
Here, note that these experiments utilize the two-mode squeezing interaction Hamiltonian (via blue detuning), which is in contrast to the direct quantum transduction that utilizes the beam-splitter interaction Hamiltonian (see Supplementary Information for a derivation of these two types of the interaction Hamiltonian).
An advantage of this quantum-teleportation based method is that we can bypass the stringent requirement of $\eta>1/2$, since the quantum capacity for this method is nonzero even under the same device condition when the quantum capacity for the direct quantum transduction is zero \cite{Wu2021}.

\section*{Summary}
To summarize, we have overviewed the theoretical basics and the experimental progress of the microwave-to-optical quantum transduction, which is an essential quantum technology for the interconnects between quantum devices operated at microwave frequencies over optical fibers.
We have overviewed the latest experimental progress of various methods for the microwave-to-optical transduction on a one-by-one basis, showing explicitly the physical origins of the interaction between microwave and optical photons or the interaction between photons and the intermediate bosonic modes.
Recent experiments have demonstrated the quantum transduction from superconducting qubit to optical photon with nearly percent-level efficiencies.
Also, several studies have achieved the so-called quantum-enabled regime of $N_{\mathrm{add}}<1$ (see Fig.~\ref{Fig-Efficiency-vs-Noise}).
Here, note that there is in practice a trade-off between a high efficiency and a low added noise.
Namely, in practice the unit transduction efficiency $\eta=1$ and the zero added noise $N_{\mathrm{add}}=0$ cannot be realized simultaneously, although a transduction efficiency of $\eta>1/2$ and a low added noise of $N_{\mathrm{add}}\ll 1$ can be achieved simultaneously (see Fig.~\ref{Fig-Trade-off}).
A recent study has also demonstrated coherent optical control of a superconducting qubit via a quantum transducer with a percent-level efficiency.
In light of these advances, as illustrated in Fig.~\ref{Fig-Qubit-to-Optical}(d), a microwave-to-optical quantum transducer may enable all-optical qubit control and readout, reducing the number of microwave cables and simplifying the structure of dilution refrigerators.
Reaching the region with a high transduction efficiency of $\eta>1/2$ and a low added noise of $N_{\mathrm{add}}\ll 1$, where the quantum state transfer between distant superconducting qubits over optical fibers is enabled, is quite challenging but is expected to become possible in the near future.

\section*{Data availability}
No datasets were generated or analysed during the current study.

\section*{Acknowledgements}
We would like to thank Shintaro Sato, Mari Ohfuchi, and Norinao Kouma for their advice and support.

\section*{Author contributions}
A.S. conceived the idea, initiated the project, and wrote the manuscript.
Y.D. supervised the project.
R.M. and Y.D. contributed to the discussion, review, and revision of the manuscript at all stages.

\section*{Funding}
This study received no funding.

\section*{Competing interests}
The authors declare no competing interests.


\setcounter{figure}{0}
\setcounter{equation}{0}
\setcounter{table}{0}
\renewcommand{\thefigure}{S\arabic{figure}}
\renewcommand{\theequation}{S\arabic{equation}}
\renewcommand{\thetable}{S\Roman{table}}

\begin{widetext}
\vspace{2ex}
\begin{center}
\textbf{{\large Supplementary Information for
\\``Microwave-to-Optical Quantum Transduction of Photons for Quantum Interconnects''}}
\end{center}
\vspace{2ex}
\end{widetext}
\section{Effective Hamiltonian in the rotating-wave approximation}
In this section, we derive the Hamiltonian of a transducer system, following Refs.~\onlinecite{Aspelmeyer2014-SI,Barzanjeh2022-SI}.
We start with the following Hamiltonian of the system where an intermediate bosonic mode (such as phonons and magnons) is interacting with microwave cavity and optical cavity photons: 
\begin{align}
\hat{H}_{\mathrm{sys}}=&\, \hat{H}_{\mathrm{d}}+\hbar\omega_{\mathrm{e}}\hat{a}^\dag_{\mathrm{e}}\hat{a}_{\mathrm{e}}+\hbar\omega_{\mathrm{o}}\hat{a}^\dag_{\mathrm{o}}\hat{a}_{\mathrm{o}}+\hbar\omega_{\mathrm{m}}\hat{a}^\dag_{\mathrm{m}}\hat{a}_{\mathrm{m}}\nonumber\\
&+\hbar g\left(\hat{a}_{\mathrm{e}}+\hat{a}_{\mathrm{e}}^\dag\right)\left(\hat{a}_{\mathrm{m}}+\hat{a}_{\mathrm{m}}^\dag\right)+\hbar \zeta\hat{a}_{\mathrm{o}}^\dag\hat{a}_{\mathrm{o}}\left(\hat{a}_{\mathrm{m}}+\hat{a}_{\mathrm{m}}^\dag\right),
\end{align}
where $\hat{a}_{\mathrm{e}}$, $\hat{a}_{\mathrm{o}}$, and $\hat{a}_{\mathrm{m}}$ ($\omega_{\mathrm{e}}$, $\omega_{\mathrm{o}}$, and $\omega_{\mathrm{m}}$) are the annihilation operators (resonance frequencies) for the microwave mode, optical cavity mode, and intermediate bosonic mode, respectively.
The Hamiltonian $\hat{H}_{\mathrm{d}}=\hbar \mathcal{E}(\hat{a}_{\mathrm{o}}e^{i\omega_{\mathrm{p}}t}+\mathrm{H.c.})$ is the driving of the optical cavity by a coherent electromagnetic field with the frequency $\omega_{\mathrm{p}}$ and amplitude $\mathcal{E}$.

Applying the unitary transformation with $\hat{U}=e^{i\omega_{\mathrm{p}}\hat{a}_{\mathrm{o}}^\dag\hat{a}_{\mathrm{o}} t}$, we move to a frame rotating at the driving frequency $\omega_{\mathrm{p}}$.
The Hamiltonian in this frame is obtained by $\hat{U}\hat{H}_{\mathrm{sys}}\hat{U}^\dag-i\hbar\hat{U}(\partial \hat{U}^\dag/\partial t)$, yielding
\begin{align}
\hat{H}_{\mathrm{sys}}=&\, \hat{H}'_{\mathrm{d}}+\hbar\omega_{\mathrm{e}}\hat{a}^\dag_{\mathrm{e}}\hat{a}_{\mathrm{e}}-\hbar\delta\omega_{\mathrm{o}}\hat{a}^\dag_{\mathrm{o}}\hat{a}_{\mathrm{o}}+\hbar\omega_{\mathrm{m}}\hat{a}^\dag_{\mathrm{m}}\hat{a}_{\mathrm{m}}\nonumber\\
&+\hbar g\left(\hat{a}_{\mathrm{e}}+\hat{a}_{\mathrm{e}}^\dag\right)\left(\hat{a}_{\mathrm{m}}+\hat{a}_{\mathrm{m}}^\dag\right)+\hbar \zeta\hat{a}_{\mathrm{o}}^\dag\hat{a}_{\mathrm{o}}\left(\hat{a}_{\mathrm{m}}+\hat{a}_{\mathrm{m}}^\dag\right),
\label{Transformed-Hamiltonian}
\end{align}
where $\delta\omega_{\mathrm{o}}=\omega_{\mathrm{p}}-\omega_{\mathrm{o}}$ is the detuning of the optical cavity frequency from the driving frequency and $\hat{H}'_{\mathrm{d}}=\hbar \mathcal{E}(\hat{a}_{\mathrm{o}}+\hat{a}_{\mathrm{o}}^\dag)$.
Here, we have used that $\hat{U}\hat{a}_{\mathrm{o}}\hat{U}^\dag=e^{-i\omega_{\mathrm{p}}t}\hat{a}_{\mathrm{o}}$ and $\hat{U}\hat{a}_{\mathrm{o}}^\dag\hat{U}^\dag=e^{+i\omega_{\mathrm{p}}t}\hat{a}_{\mathrm{o}}^\dag$.
The detuning of $\delta\omega_{\mathrm{o}}<0$ ($\delta\omega_{\mathrm{o}}>0$) is usually called red (blue) detuning.
The quadratic term $\hat{a}_{\mathrm{o}}^\dag\hat{a}_{\mathrm{o}}$ can be linearized by rewriting $\hat{a}_{\mathrm{o}}$ as $\hat{a}_{\mathrm{o}}=\bar{\alpha}+\hat{a}'_{\mathrm{o}}$, where $\bar{\alpha}=\langle\hat{a}_{\mathrm{o}}\rangle$ is the average coherent amplitude and $\hat{a}'_{\mathrm{o}}$ is the fluctuating field.
The average intra-cavity photon number is given by $\bar{n}_{\mathrm{cav}}=|\bar{\alpha}|^2=|\mathcal{E}|^2/(\delta\omega_{\mathrm{o}}^2+\kappa_{\mathrm{o}}^2/4)$ with $\kappa_{\mathrm{o}}$ being the total decay rate of the optical cavity.
We then obtain the linearized Hamiltonian
\begin{align}
\hat{H}_{\mathrm{sys}}=&\, \hat{H}'_{\mathrm{d}}+\hbar\omega_{\mathrm{e}}\hat{a}^\dag_{\mathrm{e}}\hat{a}_{\mathrm{e}}-\hbar\delta\omega_{\mathrm{o}}\hat{a}^\dag_{\mathrm{o}}\hat{a}_{\mathrm{o}}+\hbar\omega_{\mathrm{m}}\hat{a}^\dag_{\mathrm{m}}\hat{a}_{\mathrm{m}}\nonumber\\
&+\hbar g\left(\hat{a}_{\mathrm{e}}+\hat{a}_{\mathrm{e}}^\dag\right)\left(\hat{a}_{\mathrm{m}}+\hat{a}_{\mathrm{m}}^\dag\right)+\hbar \zeta'\left(\hat{a}_{\mathrm{o}}+\hat{a}_{\mathrm{o}}^{\dag}\right)\left(\hat{a}_{\mathrm{m}}+\hat{a}_{\mathrm{m}}^\dag\right),
\label{Transformed-Hamiltonian2}
\end{align}
where $\zeta'=\zeta|\bar{\alpha}|$ and we have replaced $\hat{a}'_{\mathrm{o}}$ by $\hat{a}_{\mathrm{o}}$ for simplicity.

We further apply the unitary transformation $\hat{U}=\exp[i(\omega_{\mathrm{e}}\hat{a}^\dag_{\mathrm{e}}\hat{a}_{\mathrm{e}}-\delta\omega_{\mathrm{o}}\hat{a}^\dag_{\mathrm{o}}\hat{a}_{\mathrm{o}}+\omega_{\mathrm{m}}\hat{a}^\dag_{\mathrm{m}}\hat{a}_{\mathrm{m}})t]$ to Eq.~(\ref{Transformed-Hamiltonian2}).
Keeping only the relevant terms, we obtain
\begin{align}
\hat{H}_{\mathrm{sys}}=&\, \hbar g\left(\hat{a}_{\mathrm{e}}e^{-i\omega_{\mathrm{e}}t}+\hat{a}_{\mathrm{e}}^\dag e^{i\omega_{\mathrm{e}}t}\right)\left(\hat{a}_{\mathrm{m}}e^{-i\omega_{\mathrm{m}}t}+\hat{a}_{\mathrm{m}}^\dag e^{i\omega_{\mathrm{m}}t}\right)\nonumber\\
&+\hbar \zeta'\left(\hat{a}_{\mathrm{o}}e^{i\delta\omega_{\mathrm{o}}t}+\hat{a}_{\mathrm{o}}^{\dag}e^{-i\delta\omega_{\mathrm{o}}t}\right)\left(\hat{a}_{\mathrm{m}}e^{-i\omega_{\mathrm{m}}t}+\hat{a}_{\mathrm{m}}^\dag e^{i\omega_{\mathrm{m}}t}\right).
\label{Transformed-Hamiltonian3}
\end{align}
When the system is under the resonance condition such that $\omega_{\mathrm{e}}=-\delta\omega_{\mathrm{o}}=\omega_{\mathrm{m}}$ and in the resolved sideband regime where $4\omega_{\mathrm{m}}\gg \kappa_{\mathrm{e}},\kappa_{\mathrm{o}}$, only the resonant terms are kept while the rapidly oscillating terms at $\pm 2\omega_{\mathrm{m}}$ can be neglected (i.e., the rotating-wave approximation can be applied).
In this case, the interaction Hamiltonian of beam-splitter type is relevant:
\begin{align}
\hat{H}_{\mathrm{sys}}=\hbar g\left(\hat{a}_{\mathrm{e}}\hat{a}_{\mathrm{m}}^\dag+\hat{a}_{\mathrm{e}}^\dag\hat{a}_{\mathrm{m}}\right)+\hbar \zeta'\left(\hat{a}_{\mathrm{o}}\hat{a}_{\mathrm{m}}^\dag+\hat{a}_{\mathrm{o}}^\dag\hat{a}_{\mathrm{m}}\right),
\end{align}
of which form is used throughout the main text.

On the other hand, when the pump frequency is blue detuned such that $\omega_{\mathrm{e}}=\delta\omega_{\mathrm{o}}=\omega_{\mathrm{m}}$, the interaction between optical photons and the intermediate bosonic mode takes the form of two-mode squeezing type (i.e., parametric downconversion type):
\begin{align}
\hat{H}_{\mathrm{sys}}=\hbar g\left(\hat{a}_{\mathrm{e}}\hat{a}_{\mathrm{m}}^\dag+\hat{a}_{\mathrm{e}}^\dag\hat{a}_{\mathrm{m}}\right)+\hbar \zeta'\left(\hat{a}_{\mathrm{o}}^\dag\hat{a}_{\mathrm{m}}^\dag+\hat{a}_{\mathrm{o}}\hat{a}_{\mathrm{m}}\right),
\end{align}
which generates entanglement between optical photons and the intermediate bosonic mode.

\section{Input-output formalism \label{Appendix-Input-output}}
In this section, we briefly review the input-output formalism, following Ref.~\onlinecite{Clerk2010-SI}.
First, in what follows let us for concreteness derive the equation of motion for the microwave cavity mode, which constitutes Eq.~(\ref{equation-of-motion-generic-form}) in the main text, 
\begin{align}
\dot{\hat{a}}_{\mathrm{e}}=\frac{i}{\hbar}[\hat{H}_{\mathrm{sys}},\hat{a}_{\mathrm{e}}]-\frac{\kappa_{\mathrm{e}}}{2}\hat{a}_{\mathrm{e}}-\sqrt{\kappa_{\mathrm{e,e}}}\hat{a}_{\mathrm{e,in}}-\sqrt{\kappa_{\mathrm{e,i}}}\hat{a}_{\mathrm{e,th}},
\label{EOM-microwave-cavity-mode}
\end{align}
where $\hat{a}_{\mathrm{e,in}}$ and $\hat{a}_{\mathrm{e,th}}$ are the input mode operators for itinerant microwave photons and thermal (noise) photons, respectively, and $\kappa_{\mathrm{e}}=\kappa_{\mathrm{e,e}}+\kappa_{\mathrm{e,i}}$.
The equations of motion for the intermediate bosonic mode $\hat{a}_{\mathrm{m}}$ and the optical cavity mode  $\hat{a}_{\mathrm{o}}$ can also be derived in a similar way.

The total Hamiltonian describing the itinerant microwave photons $\hat{b}_p$ and the microwave cavity mode $\hat{a}_{\mathrm{e}}$ reads
\begin{align}
\hat{H}_{\mathrm{total}}=\hat{H}_{\mathrm{sys}}+\sum_{p}\hbar\omega_{b,p}\hat{b}_p^\dag\hat{b}_p-i\hbar \sum_{p}\left(f_{b,p}\hat{a}^\dag_{\mathrm{e}}\hat{b}_{p}-f^*_{b,p}\hat{b}_{p}^\dag\hat{a}_{\mathrm{e}}\right),
\label{Total-Hamiltonian-with-bath}
\end{align}
where $p$ denotes the quantum numbers for the independent modes satisfying $[\hat{b}_p,\hat{b}_{p'}^\dag]=\delta_{p,p'}$.
We have introduced a constant coupling strength $f_b$ within the Markov approximation.

The Heisenberg equations of motion for the itinerant microwave photons and the microwave cavity mode read, respectively, $\dot{\hat{a}}_{\mathrm{e}}=(i/\hbar)[\hat{H}_{\mathrm{total}},\hat{a}_{\mathrm{e}}]$ and $\dot{\hat{b}}_p=(i/\hbar)[\hat{H}_{\mathrm{total}},\hat{b}_p]$.
Explicitly, we have
\begin{align}
\dot{\hat{a}}_{\mathrm{e}}=\frac{i}{\hbar}[\hat{H}_{\mathrm{sys}},\hat{a}_{\mathrm{e}}]-\sum_p f_{b,p}\hat{b}_p,
\label{EOM-for-a_e}
\end{align}
and
\begin{align}
\dot{\hat{b}}_p=-i\omega_p\hat{b}_p+f^*_{b,p}\hat{a}_{\mathrm{e}}.
\label{EOM-for-b}
\end{align}
The equation of motion for $\hat{b}_p$ [Eq.~(\ref{EOM-for-b})] can be solved exactly, yielding for $t>t_0$
\begin{align}
\hat{b}_p(t)=e^{-i\omega_p(t-t_0)}\hat{b}_p(t_0)+\int_{t_0}^t dt' f^*_{b,p} e^{-i\omega_p(t-t')}\hat{a}_{\mathrm{e}}(t').
\label{Solution-of-b}
\end{align}
Using the Fermi's golden rule, the transition rate between the itinerant microwave photons and the microwave cavity mode is defined by
\begin{align}
\kappa_{\mathrm{e,e}}(\omega_{\mathrm{e}})=2\pi\sum_p |f_{b,p}|^2\delta(\omega_{\mathrm{e}}-\omega_{b,p}).
\label{Fermis-golden-rule}
\end{align}
Substituting Eq.~(\ref{Solution-of-b}) into Eq.~(\ref{EOM-for-a_e}) with the use of Eq.~(\ref{Fermis-golden-rule}) and making the Markov approximation under which we can set $\kappa_{\mathrm{e,e}}(\omega_{\mathrm{e}})=\kappa_{\mathrm{e,e}}=\textrm{const.}$, we obtain \cite{Clerk2010-SI}
\begin{align}
\dot{\hat{a}}_{\mathrm{e}}=\frac{i}{\hbar}[\hat{H}_{\mathrm{sys}},\hat{a}_{\mathrm{e}}]-\frac{\kappa_{\mathrm{e,e}}}{2}\hat{a}_{\mathrm{e}}-\sum_p f_{b,p}e^{-i\omega_p(t-t_0)}\hat{b}_p(t_0).
\end{align}
Here, let us introduce the input mode operator defined by
\begin{align}
\hat{a}_{\mathrm{e,in}}(t)=\frac{1}{\sqrt{2\pi\rho_b}}\sum_p e^{-i\omega_p(t-t_0)}\hat{b}_p(t_0),
\label{Input-mode-operator}
\end{align}
where $t>t_0$ and $\rho_b=\sum_p\delta(\omega_{\mathrm{e}}-\omega_{b,p})$ is the density of states.
We also assume that the coupling strength $f_{b,p}$ is constant such that $\sqrt{|f_{b,p}|^2}\equiv f$, which gives rise to the simplified Fermi's golden rule $\kappa_{\mathrm{e,e}}=2\pi f^2 \rho_b$.
Finally, using Eq.~(\ref{Input-mode-operator}), we arrive at the equation of motion for the microwave cavity mode interacting with itinerant microwave photons:
\begin{align}
\dot{\hat{a}}_{\mathrm{e}}=\frac{i}{\hbar}[\hat{H}_{\mathrm{sys}},\hat{a}_{\mathrm{e}}]-\frac{\kappa_{\mathrm{e,e}}}{2}\hat{a}_{\mathrm{e}}-\sqrt{\kappa_{\mathrm{e,e}}}\hat{a}_{\mathrm{e,in}},
\label{EOM-with-contribution-from-itinerant-microwave}
\end{align}
which takes a well-established form.

Similarly, we can add the bath Hamiltonian for the thermal (noise) photons $\hat{c}_q$ and the interaction Hamiltonian between the microwave cavity mode and the thermal photons $\hat{H}_{\mathrm{int}}=\sum_{q}\hbar\omega_{c,q}\hat{c}_q^\dag\hat{c}_q-i\hbar \sum_{q}\left(f_{c,q}\hat{a}^\dag_{\mathrm{e}}\hat{c}_{q}-f^*_{c,q}\hat{c}_{q}^\dag\hat{a}_{\mathrm{e}}\right)$ to the total Hamiltonian~(\ref{Total-Hamiltonian-with-bath}).
Then, we can calculate the contribution from the thermal photons to the equation of motion for the microwave cavity mode in the same way as deriving Eq.~(\ref{EOM-with-contribution-from-itinerant-microwave}).
It gives rise to the additional term $-(\kappa_{\mathrm{e,i}}/2)\hat{a}_{\mathrm{e}}-\sqrt{\kappa_{\mathrm{e,i}}}\hat{a}_{\mathrm{e,th}}$ on the right-hand side of Eq.~(\ref{EOM-with-contribution-from-itinerant-microwave}).
In the end, we obtain Eq.~(\ref{EOM-microwave-cavity-mode}).

Next, let us consider the relation between the input and output modes, as represented by Eq.~(\ref{input-output-generic-form}) in the main text.
To this end, notice that there is another solution to the equation of motion for $\hat{b}_p$ [Eq.~(\ref{EOM-for-b})] such that
\begin{align}
\hat{b}_p(t)=e^{-i\omega_p(t-t_1)}\hat{b}_p(t_1)-\int_{t}^{t_1} dt' f^*_{b,p} e^{-i\omega_p(t-t')}\hat{a}_{\mathrm{e}}(t'),
\label{Solution-of-b}
\end{align}
where $t<t_1$.
Then, defining the output mode operator 
$\hat{a}_{\mathrm{e,out}}(t)=\frac{1}{\sqrt{2\pi\rho_b}}\sum_p e^{-i\omega_p(t-t_1)}\hat{b}_p(t_1)$ (with $t<t_1$),
we obtain the equation of motion for the microwave cavity mode,
\begin{align}
\dot{\hat{a}}_{\mathrm{e}}=\frac{i}{\hbar}[\hat{H}_{\mathrm{sys}},\hat{a}_{\mathrm{e}}]+\frac{\kappa_{\mathrm{e,e}}}{2}\hat{a}_{\mathrm{e}}-\sqrt{\kappa_{\mathrm{e,e}}}\hat{a}_{\mathrm{e,out}}.
\label{EOM-with-contribution-from-itinerant-microwave2}
\end{align}
Subtracting Eq.~(\ref{EOM-with-contribution-from-itinerant-microwave2}) from Eq.~(\ref{EOM-with-contribution-from-itinerant-microwave}), we obtain the well-known input-output relation,
\begin{align}
\hat{a}_{\mathrm{e,out}}(t)=\hat{a}_{\mathrm{e,in}}(t)+\sqrt{\kappa_{\mathrm{e,e}}}\hat{a}_{\mathrm{e}}(t).
\label{Input-output-microwave}
\end{align}
Here, note the dimension difference between $\hat{a}_{\mathrm{e}}(t)$ and $\hat{a}_{\mathrm{e,in/out}}(t)$.
Namely, the dimension of $\hat{a}_{\mathrm{e,in/out}}(t)$ is $\mathrm{T}^{-1/2}$, while the dimension of $\hat{a}_{\mathrm{e}}(t)$ is $\mathrm{T}^0$.

\section{Transduction efficiency and added noise in the zero-stage transduction \label{Appendix-Noise}}
In this section, we derive expressions for the transduction efficiency and the added noise in the zero-stage transduction such as the one using the electro-optic effect.
In the case of zero-stage transduction, we define the vectors $\vec{c}=[\hat{a}_{\mathrm{e}},\hat{a}_{\mathrm{o}}]^T$ and $\vec{c}_{\mathrm{in}}=[\hat{a}_{\mathrm{e,in}}, \hat{a}_{\mathrm{e,th}}, \hat{a}_{\mathrm{o,in}}, \hat{a}_{\mathrm{o,th}}]^T$.
Then, the scattering matrix is written as $S^{(0)}(\omega)=I_{4}-[B^{(0)}]^T[-i\omega I_{2}+A^{(0)}]^{-1}B^{(0)}$, where $I_{4}$ ($I_2$) is the $4\times 4$ ($2\times 2$) identity matrix, $A^{(0)}$ is a $2\times 2$ matrix, and $B^{(0)}$ is a $2\times 4$ matrix.
Here,  the matrices $A$ and $B$ are given explicitly as
\begin{align}
A^{(0)}=
\begin{bmatrix}
i\omega_{\mathrm{e}}+\kappa_{\mathrm{e}}/2 & iG_{\mathrm{eo}}\\
iG_{\mathrm{eo}} & -i\delta\omega_{\mathrm{o}}+\kappa_{\mathrm{o}}/2\\
\end{bmatrix}
\end{align}
and
\begin{align}
B^{(0)}=
\begin{bmatrix}
\sqrt{\kappa_{\mathrm{e,e}}} & \sqrt{\kappa_{\mathrm{e,i}}} & 0 & 0\\
0 & 0 & \sqrt{\kappa_{\mathrm{o,e}}} & \sqrt{\kappa_{\mathrm{o,i}}}
\end{bmatrix},
\end{align}
respectively.

Introducing the susceptibilities $\chi_{\mathrm{e}}=[-i(\omega-\omega_{\mathrm{e}})+\kappa_{\mathrm{e}}/2]^{-1}$ and $\chi_{\mathrm{o}}=[-i(\omega+\delta\omega_{\mathrm{o}})+\kappa_{\mathrm{o}}/2]^{-1}$, the microwave-to-optical transduction efficiency is defined by
\begin{align}
\eta^{(0)}(\omega)&=|S^{(0)}_{3,1}(\omega)|^2\nonumber\\
&=\left|\frac{G_{\mathrm{eo}}\sqrt{\kappa_{\mathrm{e,e}}}\sqrt{\kappa_{\mathrm{o,e}}}}{\chi_{\mathrm{e}}^{-1}(\omega)\chi_{\mathrm{o}}^{-1}(\omega)+G_{\mathrm{eo}}^2}\right|^2.
\label{efficiency-definition-zero-stage-SI}
\end{align}
Similarly, the transduction efficiency of the optical-to-microwave quantum transduction is given by $|S^{(0)}_{1,3}|^2$.
In our model, where the interaction Hamiltonian is of beam-splitter type, it turns out that $|S^{(0)}_{1,3}|^2=|S^{(0)}_{3,1}|^2=\eta^{(0)}$.
Under the resonance condition $\omega=-\delta\omega_{\mathrm{o}}=\omega_{\mathrm{e}}$, we obtain an explicit expression for the transduction efficiency in terms of the cooperativity,
\begin{align}
\eta^{(0)}=\eta_{\mathrm{e}}\eta_{\mathrm{o}}\frac{4C_{\mathrm{eo}}}{(1+C_{\mathrm{eo}})^2},
\label{Transduction-efficiency-zero-stage-transduction-SI}
\end{align}
where $\eta_{\mathrm{o}}=\kappa_{\mathrm{o,e}}/\kappa_{\mathrm{o}}$, $\eta_{\mathrm{e}}=\kappa_{\mathrm{e,e}}/\kappa_{\mathrm{e}}$, and $C_{\mathrm{eo}}=4G_{\mathrm{eo}}^2/(\kappa_{\mathrm{e}}\kappa_{\mathrm{o}})$ is the cooperativity between microwave and optical photons.

The noise input operators in Eq.~(\ref{Input-output-with-noise}) in the main text are given by
$\hat{d}^{(0)}_{\mathrm{in}}=S^{(0)}_{3,1}\delta\hat{a}_{\mathrm{e,in}}+S^{(0)}_{3,2}\hat{a}_{\mathrm{e,th}}+S^{(0)}_{3,3}\hat{a}_{\mathrm{o,in}}+S^{(0)}_{3,4}\hat{a}_{\mathrm{o,th}}$ and 
$\hat{e}^{(0)}_{\mathrm{in}}=S^{(0)}_{1,1}\delta\hat{a}_{\mathrm{e,in}}+S^{(0)}_{1,2}\hat{a}_{\mathrm{e,th}}+S^{(0)}_{1,4}\hat{a}_{\mathrm{o,th}}$.
As in the case of the one-stage transduction, we can safely ignore the thermal noise of the optical cavity, i.e., $N_{\mathrm{o,th}}\approx 0$, as well as the thermal noise of the optical fiber, i.e., $N_{\mathrm{fiber}}\approx 0$ even at room temperature.
The average numbers of the input thermal noise photons are therefore obtained as
\begin{align}
N^{(0)}_{\mathrm{o,out}}=|S^{(0)}_{3,1}|^2N_{\mathrm{wg}}+|S^{(0)}_{3,2}|^2N_{\mathrm{e,th}},
\end{align}
\begin{align}
N^{(0)}_{\mathrm{e,out}}=|S^{(0)}_{1,1}|^2N_{\mathrm{wg}}+|S^{(0)}_{1,2}|^2N_{\mathrm{e,th}},
\end{align}
where $N_{\mathrm{e,th}}=(e^{\hbar\omega_{\mathrm{e}}/k_{\mathrm{B}}T_{\mathrm{e}}}-1)^{-1}$ and $N_{\mathrm{wg}}=(e^{\hbar\omega/k_{\mathrm{B}}T_{\mathrm{wg}}}-1)^{-1}$ are the Bose distribution function at temperatures $T_{\mathrm{e}}$ and $T_{\mathrm{wg}}$, respectively.

Under the resonance condition $\omega=-\delta\omega_{\mathrm{o}}=\omega_{\mathrm{e}}$, the matrix elements are given explicitly as $|S^{(0)}_{3,1}|^2=\eta=\eta_{\mathrm{e}}\eta_{\mathrm{o}}\frac{4C_{\mathrm{eo}}}{(1+C_{\mathrm{eo}})^2}$, 
$|S^{(0)}_{3,2}|^2=(1-\eta_{\mathrm{e}})\eta_{\mathrm{o}}\frac{4C_{\mathrm{eo}}}{(1+C_{\mathrm{eo}})^2}$, 
$|S^{(0)}_{1,1}|^2=\big|1-\frac{2\eta_{\mathrm{e}}}{1+C_{\mathrm{eo}}}\big|^2$, 
and $|S^{(0)}_{1,2}|^2=\eta_{\mathrm{e}}(1-\eta_{\mathrm{e}})\frac{4}{(1+C_{\mathrm{eo}})^2}$.
The added noises $N^{(0)}_{\mathrm{add,\, M\to O}}= N^{(0)}_{\mathrm{o,out}}/\eta$ for the microwave-to-optical transduction and $N^{(0)}_{\mathrm{add,\, O\to M}}= N^{(0)}_{\mathrm{e,out}}/\eta$ for the optical-to-microwave transduction are then obtained as
\begin{align}
N^{(0)}_{\mathrm{add,\, M\to O}}=N_{\mathrm{wg}}+\left(\frac{1}{\eta_{\mathrm{e}}}-1\right)N_{\mathrm{e,th}},
\label{Added-noise-optical-out-zerostage-SI}
\end{align}
\begin{align}
N^{(0)}_{\mathrm{add,\, O\to M}}=\frac{1}{\eta_{\mathrm{e}}\eta_{\mathrm{o}}}\frac{\left|1-2\eta_{\mathrm{e}}+C_{\mathrm{eo}}\right|^2}{4C_{\mathrm{eo}}}N_{\mathrm{wg}}+\frac{1-\eta_{\mathrm{e}}}{\eta_{\mathrm{o}}}\frac{1}{C_{\mathrm{eo}}}N_{\mathrm{e,th}}.
\label{Added-noise-microwave-out-zerostage-SI}
\end{align}
We see from Eqs.~(\ref{Added-noise-optical-out-zerostage-SI}) and (\ref{Added-noise-microwave-out-zerostage-SI}) that low added noises are realized by a highly over-coupled microwave port $\eta_{\mathrm{e}}\to 1$.

\end{document}